\documentclass[12pt,a4paper]{article}
\usepackage{epsfig}
\renewcommand{\thefootnote}{\fnsymbol{footnote}}
 
\def\eq#1{{Eq.~(\ref{#1})}}
\def\fig#1{{Fig.~\ref{#1}}}

\newcommand{\be}{\begin{equation}}
\newcommand{\ee}{\end{equation}}
\newcommand{\beqar}[1]{\begin{eqnarray}\label{#1}}
\newcommand{\eeqar}{\end{eqnarray}}
\newcommand{\ben}{\begin{eqnarray*}}
\newcommand{\een}{\end{eqnarray*}}

\newcommand{\as}{\alpha_s}
\newcommand{\un}{\underline}

\newcommand{\stackeven}[2]{{{}_{\displaystyle{#1}}\atop\displaystyle{#2}}}
\newcommand{\lsim}{\stackeven{<}{\sim}}
\newcommand{\gsim}{\stackeven{>}{\sim}}
\begin{document}
\title {{\bf Can Thermalization in Heavy Ion Collisions~\vspace*{10mm} be 
Described by QCD Diagrams? 
\\[1cm] }}
\author{{\bf Yuri V. Kovchegov\thanks{e-mail: yuri@mps.ohio-state.edu}}~\\[10mm]
{\it\small Department of Physics, The Ohio State University}\\
{\it\small Columbus, OH 43210} }
 
\date{March, 2005}
\maketitle
\thispagestyle{empty}
 
\begin{abstract}
The onset of thermalization in heavy ion collisions in the weak
coupling framework can be viewed as a transition from the initial
state Color Glass Condensate dynamics, characterized by the energy
density scaling like $\epsilon \sim 1/\tau$ with $\tau$ the proper
time, to the hydrodynamics-driven expansion of the quark-gluon plasma
with $\epsilon \sim 1/\tau^{4/3}$ (or higher power of $1/\tau$ for the
boost non-invariant case). We argue that, at any order of the
perturbative expansion in the QCD coupling constant, the gluon field
generated in an ultrarelativistic heavy ion collision leads to energy
density scaling as $\epsilon \sim 1/\tau$ for late times $\tau \gg
1/Q_s$. Therefore it is likely that thermalization and hydrodynamic
description of the gluon system produced in heavy ion collisions can
not result from perturbative QCD diagrams at these late times. At
earlier times with $\tau \sim 1/Q_s$ the subleading corrections to
$\epsilon$ in $1/\tau$ expansion (terms scaling like $\sim
1/\tau^{1+\Delta}$ with $\Delta >0$) may become important possibly
leading to hydrodynamic-like behavior of the gluon system. Still, we
show that such corrections do not contribute to the particle
production cross section, and are likely to be irrelevant for physical
observables. We generalize our results by including massless quarks
into the system. Thus, it appears that the apparent thermalization of
quarks and gluons, leading to success of Bjorken hydrodynamics in
describing heavy ion collisions at RHIC, can only be attributed to the
non-perturbative QCD effects.
\end{abstract}
\thispagestyle{empty}

\newpage

\renewcommand{\thefootnote}{\arabic{footnote}}
\setcounter{footnote}{0}

\setcounter{page}{1}
%%%%%%%%%%%%%%%%%%%%%%%%%%%%%%%%%%%%%%%%%%%%%%%%%%%%%%%%%%%%%%%%%%%%%%%%%%
%
\section{Introduction}
 
Understanding how the system of quarks and gluons produced in
ultrarelativistic heavy ion collisions evolves towards thermal
equilibrium is one of the central questions in our theoretical
understanding of nuclear collisions. On the one hand, there exists a
strong experimental evidence for equilibration of the quark-gluon
system produced in a heavy ion collision at RHIC. The evidence is
based on the success of hydrodynamic models of the collisions
\cite{bj,EKR,hydro1,hydro2,HN}, indicating a collective behavior of 
the quark-gluon system, and on the discovery of jet quenching
\cite{dAtaphen,dAtaphob,dAtastar,brahms,aaphenix,aaphobos,aastar,Bj,EL,BDMPSfull,EL2,Zak,SW}, 
which demonstrated the presence of strong final state
interactions. However, hydrodynamic models of the evolution of the
quark gluon system only work well, especially for the elliptic flow
$v_2$ \cite{Ollie}, if equilibration occurs in fact at a {\sl very}
early time \cite{hydro1,hydro2}, $t \lsim 0.5 \,{\rm fm}/c$, a time
whose smallness is difficult to reconcile with current dynamical
pictures of equilibration \cite{BMSS,Mueller_eq,SS,BV,DG,MG,Wongc}.

On the other hand, a complete theoretical understanding of
thermalization is still lacking. The success of saturation/Color Glass
approach \cite{glr,mq,bm,mv,k1,jkmw,dip,bk,jimwlk,GM1} in describing
particle multiplicities \cite{KN} in heavy ion collisions and particle
spectra in deuteron--gold collisions
\cite{brahms-1,brahms-2,phenix,phobos,star,klm,kkt1,aaksw,bkw,jmnv,km,dmc,kt,braun} 
(see also \cite{kst,ag}) appears to indicate that saturation/Color
Glass formalism is valid for the initial stages of heavy ion collisions
at RHIC. Our understanding of the very early pre-equilibration stages
of the collision and their description in terms of classical gluon
fields has significantly advanced in the recent years
\cite{KMW,KR,GM,KV,K2,KNV,Lappi,MS}. Based on this saturation initial conditions, 
Baier, Mueller, Schiff and Son proposed the so-called ``bottom-up''
thermalization scenario \cite{BMSS} in which multiple $2 \rightarrow
2$, $2 \rightarrow 3$ and $3 \rightarrow 2$ rescattering processes,
the importance of which was originally underlined in \cite{Wong},
would drive the system to thermal equilibration over the time scales
of the order of $\tau_0 \sim 1/\as^{13/5} Q_s$. While the estimates in
\cite{BMSS} were mostly parametric, the numerical value of this thermalization 
time appears to be much larger than $0.5$~fm needed by hydrodynamic
simulations \cite{hydro1,hydro2}.

More recently it was argued by Arnold, Lenaghan and Moore that the
``bottom-up'' thermalization scenario could be susceptible to plasma
instabilities \cite{ALM,AL}, which were advocated previously in
\cite{Uli,Mrow,RM,SR,MMR}. Such instabilities might help facilitate the
equilibration process making the thermalization time shorter than
predicted by the ``bottom-up'' scenario. However, in \cite{AL} Arnold
and Lenaghan proved a lower bound on the thermalization time, which
happened to be surprisingly close to the ``bottom-up'' estimate. In
\cite{ALMY} it has been suggested that, while complete thermalization 
may not happen until later times, an isotropization of the produced
particle distribution in momentum space may happen much faster,
leading to generation of longitudinal pressure needed for hydrodynamic
description to work.

Here we take a different approach to the problem of
thermalization. Thermalization could be thought of as a transition
between the initial conditions, which are characterized by the energy
density scaling like $\epsilon \sim 1/\tau$, and the Bjorken
hydrodynamics, which, in case of the ideal gas equation of state has
$\epsilon \sim 1/\tau^{4/3}$ \cite{bj}. (Of course at realistic
temperatures achieved in heavy ion collisions the power of $4/3$ may
become somewhat smaller: however, it is always greater than $1$ for
hydrodynamic expansion.) Therefore it appears that corrections to the
saturation/Color Glass initial conditions
\cite{KMW,KR,GM,KV,K2,KNV} would contribute towards modifying
the $\epsilon \sim 1/\tau$ scaling to some higher power. Thus one
should be interested in Feynman diagrams which would bring in
$\tau$-dependent corrections to $\epsilon \sim 1/\tau$ scaling of the
(classical) gluon fields in the initial stages of the
collisions. Unfortunately, after examining a number of diagrams, we
noticed that while many of them introduce $\tau$-dependent corrections
to the initial conditions, such corrections are subleading and small
at large $\tau$ and do not modify $\epsilon \sim 1/\tau$ scaling at
late times. After reaching this conclusion we have constructed a
general argument proving that $\epsilon \sim 1/\tau$ scaling always
dominates at late times, both for classical fields and quantum
corrections, which we are presenting here.

The paper is structured in the following way. We begin in Section 2 by
calculating the energy density of a lowest-order non-trivial classical
gluon field from \cite{KR}. As expected the energy density of the
classical field scales as $\epsilon \sim 1/\tau$. We then continue in
Section 3 by considering the most general case of boost-invariant
gluon production, which is, indeed, not limited to classical
fields. We argue that $\epsilon \sim 1/\tau$ scaling persists to all
orders in the coupling constant $\as$, as shown in \eq{ed}. The
argument is based on a simple observation (see \eq{I1}) that
$\tau$-dependent corrections to the classical gluon field may only
come in through powers of gluon virtuality $k^2$ in momentum space
with each power of $k^2$ giving rise to a power of $1/\tau$. In order
for the on-mass shell amplitude (at $k^2 =0$) to be non-singular only
positive powers of $k^2$ are allowed: hence, the corrections come in
only as inverse extra powers of $\tau$ and are negligible at late
times. In Section 3 we generalize our results to rapidity-dependent
distributions. The $\epsilon \sim 1/\tau$ scaling does not get
modified by rapidity-dependent corrections either (see
\eq{ede}). Rapidity-dependent corrections come in as, for example, powers of $k_+$, 
which is one of light cone components of the gluon's
momentum. However, as could be seen from, say, \eq{J7}, powers of
$k_+$ do not modify the $\tau$-dependence of energy density. In
Section 4 we argue that $\epsilon \sim 1/\tau$ scaling persists even
when massless quarks are included in the problem. Therefore it appears
that perturbative thermalization can not happen in heavy ion
collisions. We conclude in Section 5 by arguing that if perturbative
thermalization is impossible, than the non-perturbative QCD effects
must be responsible for the formation of quark-gluon plasma (QGP) at
RHIC \cite{DK}. We list the non-perturbative effects which we believe may be
responsible for thermalization.

\section{Energy-Momentum Tensor of Classical Gluon Field}

\begin{figure}[b]
\begin{center}
\epsfxsize=15cm
\leavevmode
\hbox{\epsffile{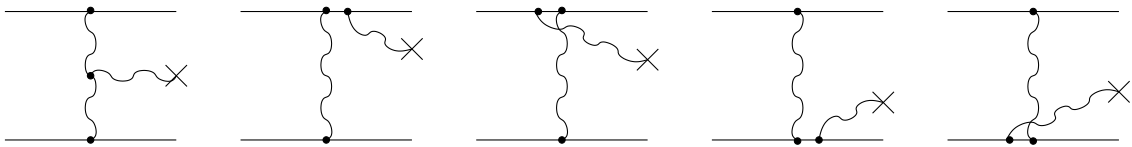}}
\end{center}
\caption{Lowest order ($\sim g^3$) gluon field produced in nuclear collisions.}
\label{twonuc}
\end{figure}

We start by calculating the energy-momentum tensor of the lowest order
gluon field produced in an ultrarelativistic heavy ion collision. This
field has been found analytically in \cite{KMW,KR} and the
corresponding Feynman diagrams are depicted here in
\fig{twonuc}. The cross in \fig{twonuc} denotes the space-time point in which we
measure the gluon field.
%%%%%%%%%%%%%%%%%%%%%%%%%%%%%
%%%%%%%%%%%%%%%%%%%%%%%%%%%%%%
The gluon field in $\partial_\mu A^\mu = 0$ covariant gauge given by
diagrams in \fig{twonuc} can be written as \cite{KR}
\be\label{lofi}
A_{\mu}^{(3) \, a} (x) = - i \int \frac{ d^4 k}{(2 \pi)^4}\, \frac{e^{- i k
\cdot x} }{k^2 + i \epsilon k_0 } \, J_\mu^{(3) \, a} (k),
\ee
with
\be\label{loj}
J_\mu^{(3) \, a} (k) \, = \, \sum_{i,j = 1}^{A_1,A_2} \, \int d^2 q \,
\frac{g^3}{ (2 \pi)^2} \, f^{abc}\, (T_i^b) \, (\tilde{T}_j^c)\, e^{i
[k_+ x_{i-} + k_- y_{j+} -
\underline{k} \cdot \underline{y}_j - \underline{q} \cdot
(\underline{x}_i - \underline{y}_j)]} \, \frac{C_\mu (k,
\underline{q})}{\underline{q}^2 (\underline{k} - \underline{q})^2}
\ee
where $C_\mu (k, \underline{q})$ is the Lipatov vertex \cite{bfkl}
\be\label{lipa}
C_\mu (k, \underline{q}) \, = \, \left( \frac{
\underline{q}^2 }{k_- + i \epsilon} - k_+ \, , \, - \frac{
(\underline{k} - \underline{q})^2 }{k_+ + i \epsilon} + k_- \, , \,
2\underline{q} - \underline{k}
\right).
\ee
The field of \eq{lofi} is given for a collision of a quark $i$ in one
of the nuclei having transverse coordinate ${\un x}_i$ and light cone
coordinate $x_{i-}$ with a quark $j$ in the other nucleus having
transverse coordinate ${\un y}_j$ and the light cone coordinate
$y_{j+}$. The matrices $(T_i^b)$ and $(\tilde{T}_j^c)$ act in the
color spaces of the quarks $i$ and $j$ correspondingly. Indeed we sum
over all quark pairs in \eq{loj}, with valence quarks $i$ being in any
of the $A_1$ nucleons in the first nucleus and valence quarks $j$
being in any one of the $A_2$ nucleons in the second nucleus.

The energy-momentum tensor of a gluon field is given by
\be\label{tmng}
T^{\mu\nu} \, = \, - F^{a \, \mu\rho} \, F^{a \, \nu}_{\ \ \ \rho} +
\frac{1}{4} \, g^{\mu\nu} \, (F^a_{\rho\sigma})^2.
\ee
We need to calculate $T^{\mu\nu}$ averaged in the wave functions of
both nuclei \cite{k1,KR}
\be\label{tmnga}
\left< T^{\mu\nu} \right> \, = \, \left< - F^{a \, \mu\rho} \, 
F^{a \, \nu}_{\ \ \ \rho} +
\frac{1}{4} \, g^{\mu\nu} \, (F^a_{\rho\sigma})^2 \right>,
\ee
where the averaging implies integrating over all possible positions of
quarks in the nucleons and nucleons in the nuclei, and taking traces
(divided by $N_c$) in the color spaces of the quarks \cite{k1,KR}. The
averaging can be represented as
\be\label{ave}
\left< \ldots \right> \, = \, \prod_{i,j=1}^{A_1, A_2} \frac{d^2 x_i}{S_{1\perp}} 
\frac{d^2 y_j}{S_{2\perp}} \frac{d x_{i-}}{a_-}  \frac{d y_{j+}}{a_+} \, 
\frac{1}{N_c^2} \, \mbox{Tr}_i [ \mbox{Tr}_j [ \ldots ]] 
\ee
where $S_{1\perp}$ and $S_{2\perp}$ are the cross sectional areas of
the two nuclei, which we for simplicity assume to be cylindrical with
the cylinder axis pointing in the beam ($z$-) direction. $a_-$ and
$a_+$ are Lorentz-contracted nucleon sizes in the $-$ and $+$
directions correspondingly, which are very small, making averaging
over $x_{i-}$ and $y_{j+}$ equivalent to just putting $x_{i-} = 0$ and
$y_{j+} =0$ \cite{KR}.

We are interested in calculating $T_{\mu\nu}$ for the gluon field
produced in a central nuclear collisions in the forward light
cone. Since the gluon field of \eq{lofi} is $o(g^3)$, we may only use
it to compute $o(g^6)$ contribution to the energy-momentum tensor, for
which we will need only the Abelian part of the field strength tensor
$F^a_{\mu\nu}$. To compute higher orders in $T_{\mu\nu}$ one would
also need higher orders in $A_\mu^a$. Using the field (\ref{lofi}) in
the Abelian part of \eq{tmnga}, performing the averaging defined in
\eq{ave} and remembering that the multiplicity distribution given by
the diagrams of \fig{twonuc} for gluons with transverse momentum ${\un
k}$, rapidity $y$, located at impact parameter ${\un b}$, is
\be\label{lomult}
\frac{d N}{d^2 k \, dy \, d^2 b} \, = \, \frac{8 \, \as^3 \, C_F}{\pi} \, 
\frac{A_1 \, A_2}{S_{1\perp} \, S_{2\perp}} \, \frac{1}{{\un k}^4} \, 
\ln \frac{k_T}{\Lambda}
\ee
we obtain after lengthy algebra (and dropping the averaging sign around
$T_{\mu\nu}$)
\begin{eqnarray}\label{tmncl}
&&T_{++} \, = \, \left( \frac{x_+}{\tau} \right)^2 \, \pi \, \int d^2 k \, \frac{d N}{d^2 k \, d \eta \, d^2 b} \ k_T^2 \, \left[ J_1 (k_T \tau) \right]^2 \nonumber ~\\
&&T_{--} \, = \, \left( \frac{x_-}{\tau} \right)^2 \, \pi \, \int d^2 k \, \frac{d N}{d^2 k \, d \eta \, d^2 b} \ k_T^2 \, \left[ J_1 (k_T \tau) \right]^2 \nonumber ~\\
&&T_{+-} \, = \, \frac{x_+ \, x_-}{\tau^2} \, \pi \, \int d^2 k \, \frac{d N}{d^2 k \, d \eta \, d^2 b} \ k_T^2 \, \left[ J_0 (k_T \tau) \right]^2 \nonumber ~\\
&&T_{ij} \, = \, \delta_{ij} \, \frac{\pi}{2} \, \int d^2 k \, \frac{d N}{d^2 k \, d \eta \, d^2 b} \ k_T^2 \, \left[ J_0 (k_T \tau) \right]^2, \ \ \ T_{+i} \, = \, T_{-i} \, = \, 0.
\end{eqnarray}
Here we used $x_\pm = (t \pm z)/\sqrt{2}$, $\tau = \sqrt{t^2 - z^2} =
\sqrt{2 x_+ x_-}$, and $k_T = |{\un k}|$. We also took advantage
of the fact that the multiplicity distribution (\ref{lomult}) is
rapidity independent and, since $T_{\mu\nu}$ should depend only on
space-time coordinates, replaced momentum space rapidity $y$ with the
space-time rapidity $\eta = (1/2) \ln (x_+ / x_-)$. (At this point
such substitution makes no difference: in Section \ref{arg2} we will
show how this substitution is formally justified in the
rapidity-dependent case.) In arriving at \eq{tmncl} we have used the
integral defined in \eq{I1} and the one given by \eq{J5} in Appendix B
with $\Delta = 0$ along with
\be
\int \frac{d^2 q}{{\un q}^2 ({\un k} - {\un q})^2} \, = \, \frac{4 \pi}{{\un k}^2} 
\, \ln \frac{k_T}{\Lambda}
\ee
where $\Lambda$ is some infrared cutoff. \eq{tmncl} is derived in the
leading logarithmic approximation in $\ln k_T/\Lambda$.

Eq. (\ref{tmncl}) gives us the energy-momentum tensor in the forward
light cone of the lowest order gluon field from \fig{twonuc} produced
in a central collision of two identical nuclei. While it is written in
a non-specific form with regards to the order of the coupling constant
$g$, we have proven Eq. (\ref{tmncl}) only at the order $o(g^6)$.

\section{Energy Density in the Boost-Invariant Approximation}

\subsection{Region of Applicability}

Let us first consider the case of high energy heavy ion collisions,
where the total rapidity interval is large enough to allow for eikonal
approximation, but not large enough for quantum BFKL-type corrections
\cite{bfkl} to become important. This is the quasi-classical regime of 
McLerran-Venugopalan model \cite{mv,k1,jkmw}. To achieve it one needs
the Bjorken $x$ variable to be small enough such that \cite{ks}
\be\label{xmv}
x \, < \, \frac{1}{2 m_N R} \, \sim \, A^{-1/3},
\ee 
which is the condition ensuring that coherent eikonal interactions are
possible in the nuclear wave functions. For corresponding rapidities,
$Y = \ln 1/x$, the condition of \eq{xmv} means that
\be\label{xmv2}
Y \, > \, \ln A^{1/3}.
\ee
Remembering that McLerran-Venugopalan model corresponds to resummation
of multiple\\ rescatterings parameter $\as^2 A^{1/3} \sim 1$ we rewrite
\eq{xmv2} as
\be\label{xmv3}
Y \, > \, \ln \frac{1}{\as^2}  \, \sim \, \ln \frac{1}{\as}.
\ee
On the other hand, the boost invariant approximation is broken down by
quantum evolution corrections, which, in the dominant leading
logarithmic approximation, bring in powers of $\as Y$
\cite{bfkl,dip,bk,jimwlk}. Indeed these corrections are negligible when $\as Y \lsim
1$, such that
\be\label{xmv4}
Y \, < \, \frac{1}{\as}.
\ee
Eqs. (\ref{xmv3}) and (\ref{xmv4}) define the rapidity interval for
nuclear collisions in which the boost invariant approximation, which
we will consider in this Section, is valid.

\subsection{Most General Boost Invariant form of $T^{\mu\nu}$}

Similar to the Bjorken approach \cite{bj} we will consider a central
collision of two very large nuclei, such that the gluon production is
translationally invariant in the transverse direction. Defining two
four-vectors in terms of light-cone coordinates
\be\label{umu}
u_\mu = \left(\frac{x_+}{\tau}, \frac{x_-}{\tau}, {\un 0} \right)
\ee
and
\be\label{vmu}
v_\mu = \left(\frac{x_+}{\tau}, - \frac{x_-}{\tau}, {\un 0} \right)
\ee
we can write the most general energy-momentum tensor for the system as
\be\label{tmn1}
T_{\mu\nu} \, = \, A(\tau) \, u_\mu u_\nu + B(\tau) \, (u_\mu v_\nu + u_\nu
v_\mu) + C(\tau) \, v_\mu v_\nu + D(\tau) \, g_{\mu\nu},
\ee
where in our convention $g_{\mu\nu} = \mbox{diag}
\{ 1,-1,-1,-1 \}$.  Here the parameters $A,B,C,D$ in \eq{tmn1} are
functions of $\tau$ only, since the large transverse extent and
azimuthal cylindrical symmetry for central collisions of the nuclei
allow us to neglect the transverse coordinate dependence, and the
assumption of boost invariance makes functions $A,B,C,D$ independent
of space-time rapidity $\eta = (1/2) \ln (x_+/x_-)$.

From \eq{tmn1} we see that
\be\label{tpp}
T_{++} \, = \, [A(\tau) + B(\tau) + C(\tau)] \left( \frac{x_+}{\tau}
\right)^2
\ee
and
\be\label{tmm}
T_{--} \, = \, [A(\tau) - B(\tau) + C(\tau)] \left( \frac{x_-}{\tau}
\right)^2. 
\ee
Due to $+ \leftrightarrow -$ symmetry of the collision of two
identical nuclei it should be possible to obtain $T_{--}$ from
$T_{++}$ after changing all $+$ indices to $-$ indices of all the
relevant four-vectors in it. This condition, when applied to
Eqs. (\ref{tpp}) and (\ref{tmm}), demands that $B(\tau) = 0$.

Rewriting the remaining non-zero functions $A, C, D$ as
\be\label{relab}
A(\tau) = \epsilon (\tau) + p (\tau), \ \ \ C(\tau) = p_3 (\tau) - p
(\tau), \ \ \ \mbox{and} \ \ \ D(\tau) = - p (\tau)
\ee
we obtain for the non-zero components of the energy momentum tensor
\begin{eqnarray}\label{tmngen}
&&T_{++} \, = \, [\epsilon (\tau) + p_3 (\tau)] \, \left( \frac{x_+}{\tau}
\right)^2, \nonumber ~\\ 
&&T_{--} \, = \, [\epsilon (\tau) + p_3 (\tau)] \, \left( 
\frac{x_-}{\tau} \right)^2, \nonumber ~\\ 
&&T_{+-} \, = \, [\epsilon (\tau) - p_3 (\tau)] \, 
\frac{x_+ x_-}{\tau^2} = [\epsilon (\tau) - p_3 (\tau)] \, \frac{1}{2}, \nonumber ~\\ 
&&T_{ij} \, = \, \delta_{ij} \, p (\tau),
\end{eqnarray}
where the indices $i,j = 1,2$ denote the transverse components of the
tensor. \eq{tmngen} gives us the most general boost-invariant
energy-momentum tensor for a collision of two very large nuclei with
the total rapidity interval satisfying conditions (\ref{xmv3}) and
(\ref{xmv4}) allowing for a boost-invariant description of the gluon
production.

At $z=0$ in the center-of-mass frame the energy-momentum tensor from
\eq{tmngen} can be written as
\begin{eqnarray}\label{tmngen0}
 T^{\mu\nu} &=&  
 \left( \matrix{ \epsilon (\tau) & 0 & 0 & 0 \cr
  0 & p (\tau) & 0 & 0 \cr
  0 & 0 & p (\tau) & 0  \cr
  0 & 0 & 0  & p_3 (\tau) \cr} \right)\, .
\end{eqnarray}
Now we can see the physical meaning of the parameterization introduced
in \eq{relab}: $\epsilon$ is the energy density and $p_3$ is the
longitudinal pressure along the beam axis ($z$-direction), which in
principle does not have to be equal to the transverse pressure
$p$. Indeed, for the case of boost-invariant Bjorken hydrodynamics
\cite{bj}, the two pressures are identical, $p_3 (\tau) = p (\tau)$. However, 
as one can show, for classical gluon fields generated in heavy ion
collisions \cite{KMW,KR,GM,KV,K2}, the longitudinal pressure component
is zero at sufficiently late times, $p_3 (\tau) = 0$, while $\epsilon
(\tau)= 2 \, p (\tau) \neq 0$ \cite{KNV}.

Applying the conservation of energy-momentum tensor condition
\be\label{cons}
\partial_\mu T^{\mu\nu} \, = \, 0
\ee
to the tensor in \eq{tmngen} we obtain
\be\label{hydroeq}
\frac{d \epsilon}{d \tau} \, = \, - \frac{\epsilon + p_3}{\tau}
\ee
similar to Bjorken hydrodynamics \cite{bj}. 

\eq{hydroeq} shows that if energy density scales with proper time as 
$\epsilon \sim 1/\tau$ then the longitudinal pressure is zero, $p_3
=0$. This is indeed the case for classical gluon production in the
initial stages of the heavy ion collisions considered in Section 2
(see also \cite{KNV}). To see this let us use the energy momentum
tensor from \eq{tmncl} in \eq{tmngen} to write
\begin{eqnarray}\label{bes}
&&\epsilon \, = \, \frac{\pi}{2} \,  \int d^2 k \, \frac{d N}{d^2 k \, d \eta \, d^2 b} \ k_T^2 \, \left\{ \left[ J_1 (k_T \tau) \right]^2 + \left[ J_0 (k_T \tau) \right]^2 \right\} \nonumber ~\\
&&p_3 \, = \, \frac{\pi}{2} \,  \int d^2 k \, \frac{d N}{d^2 k \, d \eta \, d^2 b} \ k_T^2 \, \left\{ \left[ J_1 (k_T \tau) \right]^2 - \left[ J_0 (k_T \tau) \right]^2 \right\} \nonumber ~\\
&&p \, = \, \frac{\pi}{2} \,  \int d^2 k \, \frac{d N}{d^2 k \, d \eta \, d^2 b} \ k_T^2 \, \left[ J_0 (k_T \tau) \right]^2.
\end{eqnarray}
Using the large-argument asymptotics of the Bessel functions we write
\be\label{loed}
\epsilon \bigg|_{\tau \gg 1/\langle k_T \rangle} \, \approx \, \frac{1}{\tau} \, \int d^2 k \, \frac{d N}{d^2 k \, d \eta \, d^2 b} \ k_T \, = \, \frac{1}{\tau} \, \frac{d E_T}{d \eta \, d^2 b},
\ee
which precisely agrees with the Bjorken energy density estimate
\cite{bj}. Here we assumed that the gluon spectrum is characterized by some 
typical transverse momentum $\langle k_T \rangle$, such that large
time asymptotics is defined by $\tau \gg 1/\langle k_T \rangle$. (Strictly speaking
such ``typical'' momentum for lowest order gluon field of \eq{lofi} is
the infrared cutoff $\Lambda$, but it would become the saturation
scale $Q_s \gg
\Lambda$ once multiple rescatterings are included \cite{KV,K2}.)

Similarly, using the large-argument asymptotics of the Bessel
functions one can show that
\be\label{p3}
p_3 \bigg|_{\tau \gg 1/\langle k_T \rangle} \, \approx \, 0
\ee
in agreement with \eq{loed} and \eq{hydroeq}. Thus the large time
asymptotics of the energy-momentum tensor of the lowest order
classical gluon field is given by $T_{\mu\nu} = \mbox{diag}\{\epsilon,
p, p, 0\}$ with $\epsilon = 2\, p$ and $\epsilon$ given by
\eq{loed}\footnote{It is interesting to point out that in approaching
the asymptotics of Eqs. (\ref{loed}) and (\ref{p3}) both $\epsilon$
and $p_3$ oscillate, such that $\epsilon /3$ becomes temporarily
comparable to $p_3$ at proper time $\tau \sim 1/Q_s$. While the
mathematical origin of these oscillations is clearly due to the Bessel
functions in \eq{bes}, their physical interpretation (if it exists) is
presently unclear.}. Similar results were obtained in numerical
simulations of the full classical gluon field including all orders in
multiple rescatterings \cite{KNV}.

The onset of thermalization or isotropization of the system
\cite{ALMY} should come with generation of the non-zero longitudinal
pressure $p_3$ comparable to the transverse pressure $p$. In order for
that to happen \eq{hydroeq} necessarily requires the energy density to
start scaling with $\tau$ as $\epsilon \sim 1/\tau^{1 + \Delta}$,
where $\Delta$ is some positive number. In the case of ideal Bjorken
hydrodynamics $\Delta = 1/3$.

Thus the process of thermalization in heavy ion collisions can be
viewed as a transition from the $\epsilon \sim 1/\tau$ scaling,
characteristic of free-streaming classical fields (\ref{loed}), to
$\epsilon \sim 1/\tau^{1 + \Delta}$ scaling. Below we are going to
study whether such transition can result from Feynman diagram
resummation.

\subsection{Can Boost-Invariant Bjorken Hydro Result from Feynman Diagrams?}
\label{arg1}

Let us explore what kinds of energy-momentum tensor may result from
Feynman diagram resummation. We will concentrate only on gluon fields,
and later will generalize our conclusions to include quark fields as
well. We will assume that the initial gluon field is given by the
classical field of McLerran-Venugopalan model
\cite{mv,k1,KMW,KR,GM,KV,K2,KNV}, though our results would not depend much 
on this assumption\footnote{There is a common misconception in the
community that in McLerran-Venugopalan model one assumes that $y =
\eta$: while this assumption was made in the original works on the 
subject \cite{KMW,mv}, it is actually not necessary, with all the
results of McLerran-Venugopalan model easily derivable without making
any assumptions on correlations between $\eta$ and $y$ (see
\cite{KR}).}. In the saturation scenario, the gluon fields at the
early times with $\tau \sim 1/Q_s$ are strong
\be
A^a_\mu \, \sim \, \frac{Q_s}{g}.
\ee
In calculating corresponding field strength tensor $F^a_{\mu\nu}$, one
would require both the Abelian and the non-Abelian parts of
it. However, as classical fields and their energy density (\ref{loed})
decrease with proper time, for $\tau \gsim 1/Q_s$ the Abelian part of
$F^a_{\mu\nu}$ would dominate. This is also true for quantum
corrections to classical fields. Therefore, in the following
discussion we will first restrict ourselves to calculating the Abelian
part of $T_{\mu\nu}$ only, and will later show that inclusion of
non-Abelian parts of $T_{\mu\nu}$ would not change our argument.

The most general gluon field generated through any-order Feynman
diagrams in $\partial_\mu A^\mu = 0$ covariant gauge can be written as
\be\label{ffi}
A_{\mu}^{a} (x) = - i \int \frac{ d^4 k}{(2 \pi)^4}\, \frac{e^{- i k
\cdot x} }{k^2 + i \epsilon k_0 } \, J_\mu^a (k),
\ee
where we are using the retarded regularization for the outgoing gluon
propagator $-i/k^2$ to ensure causality. The function $J_\mu^a (k)$
denotes the rest of the diagram (the truncated part). In general
$J_\mu^a (k)$ is an abbreviated notation for $J_\mu^a (k; q_1,
\lambda_1; q_2, \lambda_2; \ldots)$, which depends on momenta $q_i$ of 
extra gluons (or quarks) in the final state and on their polarizations
(helicities) $\lambda_i$. In case of the classical gluon field there
are no extra particles in the final state and momenta $q_i$'s do not
enter the expression (\ref{ffi}). (In fact, the possible particles in
the final state for classical fields can be removed by using retarded
regularization of gluon propagators \cite{Ian04}.) Quantum corrections
to the classical gluon field would inevitably bring in extra final
state particles. The resulting ``quantum'' field $A_{\mu}^{a} (x)$ in
\eq{ffi} would also depend on momenta $q_i$: again we suppress this 
dependence in the notation. Indeed, once quantum corrections are
included there is no dominant gluon field anymore: in that sense the
gluon field in \eq{ffi} is not really a field, but more like a
scattering amplitude (located on one side of the cut), with one ($k$)
of the many outgoing particle lines ($q_i$'s) being off-mass shell
with its propagator ending at a space-time point $x_\mu$.

Substituting the field from \eq{ffi} into the expression for the
energy-momentum tensor (\ref{tmnga})
\ben
\left< T^{\mu\nu} \right> \, = \, \left< - F^{a \, \mu\rho} \, 
F^{a \, \nu}_{\ \ \ \rho} +
\frac{1}{4} \, g^{\mu\nu} \, (F^a_{\rho\sigma})^2 \right>,
\een
and keeping only the Abelian parts of $F^a_{\mu\nu}$'s we obtain
\ben
T_{\mu\nu} \, = \, \int \frac{ d^4 k \, d^4 k'}{(2 \pi)^8}\,
\frac{e^{-i k \cdot x - i k' \cdot x}}{(k^2 + i \epsilon k_0) \, (k'^2
+ i \epsilon k'_0)} \, \bigg< - [k_\mu J^{a \, \rho} (k) - k^\rho
J^a_\mu (k)] \,  [k'_\nu J^a_\rho (k') - k'_\rho J^a_\nu (k')]
\een
\be\label{tmnab}
+ \frac{1}{4} \, g_{\mu\nu} \,  [k^\rho J^{a \, \sigma} (k) - k^\sigma
J^{a \, \rho} (k)] \, [k'_\rho J^a_\sigma (k') - k'_\sigma J^a_\rho
(k')]\bigg>,
\ee
where the brackets $\langle \ldots \rangle$ are defined by \eq{ave}
and now also include integration over all momenta $q_i$'s. The gluon
field in \eq{ffi} is generated by the color sources in two colliding
nuclei, which are modeled by valence quarks, just like in
McLerran-Venugopalan model \cite{mv}. Of course, the resulting gluon
field from \eq{ffi} is not necessarily classical, it includes extra
quark and gluon emissions as well as loops, just like any production
diagram with incoming valence quarks of the nuclei providing the
initial condition for the scattering process.

Since performing the transverse averaging over a very large nucleus in
the brackets on the right hand side of \eq{tmnab} puts ${\un k} = -
{\un k}'$, which results from transverse translational invariance of
gluon production, we can rewrite it as
\ben
T_{\mu\nu} \, = \, \int \frac{ d^4 k \, d^4 k'}{(2 \pi)^8}\,
\frac{e^{-i k \cdot x - i k' \cdot x}}{(k^2 + i \epsilon k_0) \, (k'^2
+ i \epsilon k'_0)} \, \bigg<\bigg< - [k_\mu J^{a \, \rho} (k) - k^\rho
J^a_\mu (k)] \,  [k'_\nu J^a_\rho (k') - k'_\rho J^a_\nu (k')]
\een
\be\label{tmnab1}
+ \frac{1}{4} \, g_{\mu\nu} \, [k^\rho J^{a \, \sigma} (k) - k^\sigma
J^{a \, \rho} (k)] \, [k'_\rho J^a_\sigma (k') - k'_\sigma J^a_\rho
(k')]\bigg>\bigg> \, \frac{(2 \pi)^2}{S_\perp} \, \delta ({\un k} + {\un
k}'),
\ee
where the double brackets $\langle\langle \ldots \rangle\rangle$
denote now the color averaging, integration over $q_i$'s, summation
over nucleons in the nuclei and averaging over longitudinal and
remaining transverse coordinates. $S_\perp$ is the cross sectional
area of the nuclei which we assume to be identical.

Let us define the following correlation function
\be\label{ddef}
D_{\mu\nu} \, \equiv \, \left<\left< J_\mu (k) \, J_\nu (k')
\right>\right>\bigg|_{{\un k} = - {\un k}'}.
\ee
Using the covariant gauge condition $\partial_\mu A^\mu = 0$, which
translates into $k_\mu \, J^\mu (k) \, = \, k'_\mu \, J^\mu (k') \, =
\, 0$, along with the $k_+ \leftrightarrow k_-$ 
(and $k'_+ \leftrightarrow k'_-$) symmetry of the collision, we can
derive the following relations between different components of
$D_{\mu\nu}$:
\begin{eqnarray}\label{drel}
D_{+i} \, = \, \frac{k_j}{2 \, k_-} \, D_{ji} \hspace*{.5cm} && \hspace*{.5cm} D_{-i} \, = \, \frac{k_j}{2 \, k_+} \, D_{ji} \nonumber \\
D_{i+} \, = \, \frac{k'_j}{2 \, k'_-} \, D_{ij} \hspace*{.5cm} && \hspace*{.5cm} D_{i-} \, = \, \frac{k'_j}{2 \, k'_+} \, D_{ij} \nonumber \\
\frac{D_{++}}{k_+ \, k'_+} \, &=& \, \frac{D_{--}}{k_- \, k'_-},
\end{eqnarray}
where the Latin indices $i, j = 1,2$ indicate the transverse
components of the correlators and two lowercase repeated Latin indices
indicate contraction over that index.

We are interested in calculating the energy density $\epsilon$ given
by (see
\eq{tmngen})
\be\label{edens}
\epsilon \, = \, \frac{1}{2} \, \left( \frac{\tau}{x_+} \right)^2 \, T_{++} + T_{+-}.
\ee
Using \eq{tmnga} we write
\be\label{tpp1}
T_{++} \, = \, - \left< F^{a \, \rho}_+ \, F^a_{+\rho} \right> \, = \,
\left< F^a_{+i} \, F^a_{+i} \right>
\ee
and
\be\label{tpm1}
T_{+-} \, = \, \frac{1}{2} \, \left< F^a_{+-} \, F^a_{+-} \right> +
\frac{1}{4} \, \left< F^a_{ij} \, F^a_{ij} \right>.
\ee
Using \eq{ffi} together with relations from \eq{drel} in
Eqs. (\ref{tpp1}) and (\ref{tpm1}) we obtain
\ben
T_{++} \, = \, \int \frac{ d^4 k \, d^4 k'}{(2 \pi)^8}\,
\frac{e^{-i k \cdot x - i k' \cdot x}}{(k^2 + i \epsilon k_0) \, (k'^2
+ i \epsilon k'_0)} \, \left[ k_+ \, k'_+ \, D_{ii} - {\un k}^2 \,
D_{++} - \frac{1}{2} \, \left( \frac{k'_+}{k_-} + \frac{k_+}{k'_-}
\right) \, k_i \, k_j \, D_{ij} \right]
\een
\be\label{tpp2}
\times \, \frac{(2 \pi)^2}{S_\perp} \, \delta ({\un k} + {\un k}')
\ee
and
\ben
T_{+-} \, = \, \int \frac{ d^4 k \, d^4 k'}{(2 \pi)^8}\,
\frac{e^{-i k \cdot x - i k' \cdot x}}{(k^2 + i \epsilon k_0) \, (k'^2
+ i \epsilon k'_0)} \, \left[ - \frac{1}{2} \, {\un k}^2 \, D_{ii} + 2
k_- \, k'_- \, D_{++} + k_i \, k_j \, D_{ij} \right]
\een
\be\label{tpm2}
\times \, \frac{(2 \pi)^2}{S_\perp} \, \delta ({\un k} + {\un k}').
\ee
Substituting Eqs. (\ref{tpp2}) and (\ref{tpm2}) into \eq{edens} yields
\ben
\epsilon \, = \, \int \frac{ d^4 k \, d^4 k'}{(2 \pi)^8}\,
\frac{e^{-i k \cdot x - i k' \cdot x}}{(k^2 + i \epsilon k_0) \, (k'^2
+ i \epsilon k'_0)} \, \left\{ \left[ \frac{1}{2} \, \left(
\frac{\tau}{x_+} \right)^2 \, k_+ \, k'_+ - \frac{1}{2} \, {\un k}^2
\right] \, D_{ii} \, + \right.
\een
\be\label{edens11}
+ \left. \left[ 2 \, k_- \, k'_- - \frac{1}{2} \, \left(
\frac{\tau}{x_+} \right)^2 \, {\un k}^2 \right] \, D_{++} + \left[ 
- \frac{1}{4} \, \left( \frac{\tau}{x_+} \right)^2 \, 
\left( \frac{k'_+}{k_-} + \frac{k_+}{k'_-} \right) \, + \, 1 \right] 
\, k_i \, k_j \, D_{ij}  \right\} \, \frac{(2 \pi)^2}{S_\perp} \, \delta ({\un k} + {\un k}').
\ee
Since the tensor structure of the correlators $D_{\mu\nu}$ from
\eq{ddef} is symmetric under $k \leftrightarrow k'$, and using the 
last relation in \eq{drel}, without any loss of generality one can
write
\be\label{f2def}
D_{++} \, = \, k_+ \, k'_+ \, f_2 (k^2, k'^2, k \cdot k', k_T),
\ee
where $f_2 (k^2, k'^2, k \cdot k', k_T)$ is some unknown
boost-invariant function, which, due to rapidity independence of the
problem, depends only on $k^2$, $k'^2$, $k \cdot k'$ and on the
magnitude of the transverse momentum $k_T$. In general, dependence of
$f_2$ on $k \cdot k'$ might lead to rapidity dependence: however, as
we will see below the resulting leading energy density is still boost
invariant. $f_2 (k^2, k'^2, k \cdot k', k_T)$ is symmetric under the
interchange $k^2 \leftrightarrow k'^2$. Similarly
\be\label{f1def}
D_{ii} \, = \, f_1 (k^2, k'^2, k \cdot k', k_T)
\ee
and
\be\label{f3def}
k_i \, k_j \, D_{ij} \, = \, f_3 (k^2, k'^2, k \cdot k', k_T)
\ee
with $f_1$ and $f_3$ also some symmetric functions under $k^2
\leftrightarrow k'^2$.  Using these redefinitions we can rewrite
\eq{edens11} as
\ben
\epsilon \, = \, \int \frac{ d^4 k \, d^4 k'}{(2 \pi)^8}\,
\frac{e^{-i k \cdot x - i k' \cdot x}}{(k^2 + i \epsilon k_0) \, (k'^2
+ i \epsilon k'_0)} \, \left\{ \left[ \frac{1}{2} \, \left(
\frac{\tau}{x_+} \right)^2 \, k_+ \, k'_+ - \frac{1}{2} \, {\un k}^2
\right] \,  f_1 (k^2, k'^2,  k \cdot k', k_T) \, + \right.
\een
\ben
+  \left[ 2 \, k_- \, k'_- - \frac{1}{2} \, \left(
\frac{\tau}{x_+} \right)^2 \, {\un k}^2 \right] \, k_+ \, k'_+ \, 
f_2 (k^2, k'^2, k \cdot k', k_T) +
\een
\be\label{edens2}
+ \left. \left[ - \frac{1}{4} \, \left( \frac{\tau}{x_+} \right)^2 \,
\left( \frac{k'_+}{k_-} + \frac{k_+}{k'_-} \right) \, + \, 1 \right] 
\, f_3 (k^2, k'^2, k \cdot k', k_T) \right\} \, \frac{(2 \pi)^2}{S_\perp} 
\, \delta ({\un k} + {\un k}').
\ee

For reasons which will become apparent in a moment, we are interested
in determining the following combination of $f$'s
\ben
f_1 (k^2=0, k'^2=0,  k \cdot k' =0, k_T) - k_T^2 f_2 (k^2=0, k'^2=0, k \cdot k' =0, k_T)
\een
\be 
- \frac{2}{k_T^2} \, f_3 (k^2=0, k'^2=0, k \cdot k' =0, k_T).
\ee
To calculate it we compare \eq{tpp2} with
\ben
T_{++} \, = \, \int \frac{ d^4 k \, d^4 k'}{(2 \pi)^8}\,
\frac{e^{-i k \cdot x - i k' \cdot x}}{(k^2 + i \epsilon k_0) \, (k'^2
+ i \epsilon k'_0)} \, \bigg<\bigg< - [k_+ J^{a \, \rho} (k) - k^\rho
J^a_+ (k)] 
\een
\be\label{f2}
\times \,  [k'_+ J^a_\rho (k') - k'_\rho J^a_+ (k')] \bigg>\bigg> 
\, \frac{(2 \pi)^2}{S_\perp} \, \delta ({\un k} + {\un k}'),
\ee
which follows from \eq{tmnab}. Equating the integrands of
Eqs. (\ref{tpp2}) and (\ref{f2}) we derive
\ben
k_+ \, k'_+ \, f_1 (k^2, k'^2, k \cdot k', k_T) 
- {\un k}^2 \, k_+ \, k'_+ \, f_2 (k^2, k'^2, k \cdot k', k_T) -
\een
\be\label{f3}
 - \frac{1}{2} \, \left( \frac{k'_+}{k_-} + \frac{k_+}{k'_-}
\right) \, f_3 (k^2, k'^2, k \cdot k', k_T) 
\, = \, \bigg<\bigg< - [k_+ J^{a \, \rho} (k) - k^\rho J^a_+ (k)] \,
[k'_+ J^a_\rho (k') - k'_\rho J^a_+ (k')] \bigg>\bigg>.
\ee
Putting $k = - k'$ and $k^2 = k'^2 =0$ in \eq{f3} and employing the
fact that in covariant gauge $k^\rho J^a_\rho (k) =0$ we obtain
\ben
f_1 (k^2=0, k'^2=0, k \cdot k' =0, k_T) - k_T^2 f_2 (k^2=0, k'^2=0, k \cdot k' =0, k_T) -
\een
\be\label{f5}
- \frac{2}{k_T^2} \, f_3 (k^2=0, k'^2=0, k \cdot k' =0, k_T) \,
= \, - \bigg\langle\bigg\langle J^{a \, \rho} (k)
\, J^a_\rho (-k) \bigg\rangle\bigg\rangle \bigg|_{k^2 = 0}.
\ee
Finally, since in order to construct the amplitude out of the field
given by \eq{ffi} one needs to truncate the field and put the outgoing
gluon's momentum on the mass shell, $k^2 = 0$, we see that $J^{a \,
\rho} (k)$ at $k^2 = 0$ is nothing but a production amplitude for a
real gluon carrying momentum $k$ (without convolution with the
polarization vector). The corresponding multiplicity distribution of
the produced gluons is given by
\be
\frac{dN}{d^2 k \, dy} \, = \, \frac{1}{2 (2 \pi)^3} \, 
\bigg\langle\bigg\langle J^{a \, \rho} (k)
\, J^a_\rho (-k) \bigg\rangle\bigg\rangle \bigg|_{k^2 = 0}.
\ee
Therefore,
\ben
f_1 (k^2=0, k'^2=0, k \cdot k' =0, k_T) - k_T^2 f_2 (k^2=0, k'^2=0, k \cdot k' =0, k_T) 
\een
\be\label{f6}
- \frac{2}{k_T^2} \, f_3 (k^2=0, k'^2=0, k \cdot k' =0, k_T) \, = \, 
- 2 (2 \pi)^3 \, \frac{dN}{d^2 k \, dy}
\ee
and, for a cylindrical nucleus,
\ben
\frac{1}{S_\perp} \, \bigg[ f_1 (k^2=0, k'^2=0, k \cdot k' =0, k_T) 
- k_T^2 f_2 (k^2=0, k'^2=0, k \cdot k' =0, k_T) 
\een
\be\label{f7}
- \frac{2}{k_T^2} \, f_3 (k^2=0, k'^2=0, k \cdot k' =0, k_T) \bigg] \,
= \, - 2 (2 \pi)^3 \, \frac{dN}{d^2 k \, dy \, d^2 b}.
\ee

Now let us get back to \eq{edens2}. Rewriting for each of the $f$'s
\ben
f_i (k^2 , k'^2 , k \cdot k', k_T) \, = \, f_i (k^2 =0, k'^2 =0, k
\cdot k' =0, k_T) + 
\een
\be\label{iter1}
+ [f_i (k^2 , k'^2 , k \cdot k', k_T) - f_i (k^2 =0, k'^2 =0, k \cdot k' =0,
k_T)]
\ee
and keeping only the $f_i (k^2 =0, k'^2 =0, k \cdot k' =0, k_T)$ in
\eq{edens2} we can perform the longitudinal momenta ($k_+, k_-, k'_+,
k'_-$) integrations with the help of \eq{J7} from Appendix B obtaining
\ben
\epsilon \, \approx \, - \frac{1}{8 \, S_\perp} \, \int \frac{d^2 k}{(2 \, \pi)^2} \, 
k_T^2 \, \left\{ \left[ J_1 (k_T \tau) \right]^2 + \left[ J_0 (k_T \tau) \right]^2 
\right\}
\een
\ben
\times \, \bigg[ f_1 (k^2=0, k'^2=0, k \cdot k' =0, k_T) - k_T^2 f_2 (k^2=0, k'^2=0, 
k \cdot k' =0, k_T)
\een 
\be\label{edens33}
- \frac{2}{k_T^2} \, f_3 (k^2=0, k'^2=0, k \cdot k' =0, k_T) \bigg],
\ee
which, after employing \eq{f7} becomes
\be\label{edens3}
\epsilon \, \approx \, \frac{\pi}{2} \,  \int d^2 k \, 
\frac{d N}{d^2 k \, d \eta \, d^2 b} 
\ k_T^2 \, \left\{ \left[ J_1 (k_T \tau) \right]^2 + \left[ J_0 (k_T \tau) \right]^2 
\right\}. 
\ee
(Again we have used the rapidity-independence of the gluon spectrum
$\frac{d N}{d^2 k \, d \eta \, d^2 b}$ to replace $y$ with $\eta$.) 

One might worry that the functions $f_i (k^2 , k'^2 , k \cdot k',
k_T)$ may not have a finite $k^2,k'^2, k\cdot k^\prime \rightarrow 0$
limit, which would be dangerous for the decomposition of
\eq{iter1} \cite{AMY2}. However, let us remind the reader that the quantity $J_\mu^a
(k)$ defined in \eq{ffi} has the meaning of (truncated) gluon
production amplitude for the off-shell gluon with virtuality $k^2$. In
the $k^2 \rightarrow 0$ limit $J_\mu^a (k)$ becomes the gluon
production amplitude for an on-shell gluon, and is indeed
finite. Therefore, the correlation functions $D_{\mu\nu}$ from
\eq{ddef}, which in the $k^2,k'^2, k\cdot k^\prime \rightarrow 0$
limit have the meaning of the gluon production amplitude squared (but
without the Lorentz index contraction), are also finite in this
limit. This implies that the functions $f_i (k^2 , k'^2 , k \cdot k',
k_T)$, defined in terms of various components of $D_{\mu\nu}$ in
Eqs. (\ref{f2def}), (\ref{f1def}) and (\ref{f3def}), are finite in the
$k^2,k'^2, k\cdot k^\prime \rightarrow 0$ limit.

For the proper time $\tau$ much larger than $1/\langle k_T \rangle$,
with $k_T$ the typical transverse momentum in the distribution
$\frac{d N}{d^2 k \, d
\eta \, d^2 b}$, \eq{edens3} becomes
\be\label{ed}
\epsilon \bigg|_{\tau \gg 1/\langle k_T \rangle} \, \approx \, \frac{1}{\tau} \, \int d^2 k \, 
\frac{d N}{d^2 k \, d \eta \, d^2 b} \ k_T \, = \, \frac{1}{\tau} \, 
\frac{d E_T}{d \eta \, d^2 b},
\ee
i.e., it falls off as $1/\tau$. 

Therefore, we have shown that the energy density $\epsilon$ of a gluon
field produced in a heavy ion collision always has a non-zero term
scaling as $\sim 1/\tau$. However, to demonstrate that this term
dominates at late times, we still need to prove that it does not get
canceled by the terms we left out in writing down the decomposition of
\eq{iter1} and keeping the first terms only. Thus we have to analyze
the contribution arising from substituting the terms from the square
brackets of \eq{iter1} into \eq{edens2}. We need to show that such
contributions fall off faster than $1/\tau$, and, therefore, can be
neglected at late times. Here we will demonstrate that this is true
for one of the terms --- the $f_1$-term in \eq{edens2}. The proof for
the other two terms on the right hand side of \eq{edens2} would be
analogous.

Substituting the square brackets from \eq{iter1} into \eq{edens2} we
obtain the following contribution to the energy density, which we have
to prove to be small:
\ben
\frac{1}{2} \, \int \frac{ d^4 k \, d k'_+ \, d k'_-}{(2 \pi)^6 \, S_\perp}\,
\frac{e^{-i k \cdot x - i k' \cdot x}}{(k^2 + i \epsilon k_0) \, (k'^2
+ i \epsilon k'_0)} \, \left[ \left( \frac{\tau}{x_+} \right)^2 \,
k_+ \, k'_+ - {\un k}^2 \right] 
\een
\be\label{contr1}
\times \,[f_1 (k^2, k'^2, k \cdot k', k_T) - f_1 (k^2 =0, k'^2 =0, k \cdot k' =0, k_T)].
\ee
For a wide range of amplitudes one can write
\ben
f_1 (k^2, k'^2, k \cdot k', k_T) - f_1 (k^2 =0, k'^2 =0, k \cdot k'
=0, k_T) \, = \, (k^2 \, k'^2 )^{\Delta_1} \, g^{(1)} (k^2, k'^2, k
\cdot k', k_T) +
\een
\be\label{iter2}
+ [(k+k')^2]^{\Delta_2} \, g^{(2)} (k^2, k'^2, k
\cdot k', k_T),
\ee
where $g^{(1)} (k^2 =0, k'^2 =0, k \cdot k' =0, k_T) \neq 0$, $g^{(2)}
(k^2 =0, k'^2 =0, k \cdot k' =0, k_T) \neq 0$, and $\Delta_1, \Delta_2 > 0$. In
arriving at \eq{iter2} we have also used the fact that $f_1 (k^2,
k'^2, k \cdot k', k_T) \, = \, f_1 (k'^2, k^2, k \cdot k', k_T)$, which follows from the $k
\leftrightarrow k'$ symmetry in the definition of $f_1 (k^2, k'^2, k \cdot k',
k_T)$ given by \eq{f1def} along with \eq{ddef}. In \eq{iter2} we put a
power of $(k+k')^2$ instead of a power of $k \cdot k'$ in front of
$g^{(2)}$. Similarly to \eq{iter1} we write
\ben
g^{(i)} (k^2 , k'^2 , k \cdot k', k_T) \, = \, g^{(i)} (k^2 =0, k'^2
=0, k \cdot k'=0, k_T) + 
\een
\be\label{iter3}
+ [g^{(i)} (k^2 , k'^2 , k \cdot k', k_T) - g^{(i)} (k^2 =0,
k'^2 =0, k \cdot k'=0, k_T)]
\ee
for $i=1,2$.  Substituting the first term on the right hand side of
\eq{iter3} for $g^{(1)}$ into the first term on the right hand side of
\eq{iter2}, and then plugging the resulting contribution into \eq{contr1} yields
\be\label{iter4}
\frac{1}{2} \, \int \frac{ d^4 k \, d k'_+ \, d k'_-}{(2 \pi)^6 \,
S_\perp}\, \frac{e^{-i k \cdot x - i k' \cdot x}}{(k^2 + i \epsilon
k_0) \, (k'^2 + i \epsilon k'_0)} \, \left\{ \left( \frac{\tau}{x_+}
\right)^2 \, k_+ \, k'_+ - {\un k}^2 \right\} \,  (k^2 \, k'^2
)^{\Delta_1} \, g^{(1)} (0, 0, 0, k_T).
\ee
Performing the $k_+, k_-, k'_+, k'_-$ integrations in \eq{iter4} with
the help of \eq{I1} in Appendix A and \eq{J5} with $\lambda =1$ in
Appendix B we obtain
\be\label{iter5}
\frac{e^{2 \pi i \Delta_1}}{8 \, S_\perp \, \Gamma (1-\Delta_1)^2} \,
\int \frac{ d^2 k}{(2 \pi)^2} \, g^{(1)} (0, 0, 0, k_T) \, k_T^2 \, 
\left( \frac{2 \, k_T}{\tau} \right)^{2 \, \Delta_1} \, \left\{ \left[ J_{-\Delta_1 -1} (k_T \tau) 
\right]^2 + \left[ J_{-\Delta_1} (k_T \tau) \right]^2 
\right\},
\ee
which, for $\tau \gg 1/\langle k_T \rangle$, scales as
\be\label{iter6}
\sim \frac{1}{\tau^{1 + 2 \Delta_1}},
\ee
and is, therefore, negligibly small at late proper times compared to
the leading contribution to energy density given by \eq{ed}. (Here we
assume that the typical transverse momentum $\langle k_T \rangle$ is
the same for $\frac{d N}{d^2 k \, d \eta \, d^2 b}$ in \eq{edens3} and
for $g^{(1)} (0, 0, k_T)$ in \eq{iter5}: both functions result from
expanding the same amplitude in powers of $k^2 \, k'^2$, and no new
scale can arise from such an expansion, which justifies our
assumption.) The particular way of regularizing the $k^2$ branch cut
used in \eq{I1} and \eq{J1} is not essential for arriving at
\eq{iter6}, since other regularizations would yield the same result.

The second term on the right hand side of \eq{iter2} gives a similarly
small contribution. To see this we substitute the first term on the
right hand side of \eq{iter3} for $g^{(2)}$ into the second term on
the right hand side of \eq{iter2}, and then substitute the result into
\eq{contr1} obtaining
\ben
\int \frac{ d^4 k \, d k'_+ \, d k'_-}{(2 \pi)^6 \, 2 \, 
S_\perp}\, \frac{e^{-i k \cdot x - i k' \cdot x}}{(k^2 + i \epsilon
k_0) \, (k'^2 + i \epsilon k'_0)} \, \left\{ \left( \frac{\tau}{x_+}
\right)^2 \, k_+ \, k'_+ - {\un k}^2 \right\} [(k+k')^2]^{\Delta_2} 
\, g^{(2)} (0, 0, 0, k_T) \, =
\een
\be\label{iter7}
= \, [- \partial_\mu \partial^\mu]^{\Delta_2} \frac{1}{8 \, S_\perp} \,
\int \frac{ d^2 k}{(2 \pi)^2} \, g^{(2)} (0, 0, 0, k_T) \, k_T^2 \, 
\, \left\{ \left[ J_{1} (k_T \tau) \right]^2 + \left[ J_{0} (k_T \tau) 
\right]^2 \right\}.
\ee
For $\tau \gg 1/\langle k_T \rangle$ the integral on the right of
\eq{iter7} scales as $\sim 1/\tau$: applying the derivatives we see that 
the whole expression in \eq{iter7} scales as
\be\label{iter8}
\sim \frac{1}{\tau^{1 + 2 \Delta_2}},
\ee
and is also negligibly small at late proper times compared to the
leading contribution to energy density given by \eq{ed}.

For the second term on the right hand side of \eq{iter3}, $g^{(i)}
(k^2, k'^2, k \cdot k', k_T) - g^{(i)} (0,0,0,k_T)$ with $i=1$(or
$i=2$), one can repeat the procedure outlined above for $f_1 (k^2,
k'^2, k \cdot k', k_T) - f_1 (0,0,0,k_T)$, using the redefinition just
like in \eq{iter2} and showing that the leading term in the resulting
decomposition, similar to \eq{iter3}, falls off faster with $\tau$
than \eq{iter6} (or \eq{iter8}). Iterating the procedure would
generate a series of corrections falling off at progressively higher
powers of $1/\tau$, all of which could be neglected at $\tau \gg
1/\langle k_T \rangle$.

Of course the assumption of \eq{iter2}, while quite general, does not
include all the possibilities. One might imagine other ways for $f_1
(k^2, k'^2, k \cdot k', k_T) - f_1 (0,0,0, k_T)$ to approach zero as
$k^2, k'^2, k \cdot k' \rightarrow 0$: it might scale as $1 / (\ln k^2
\, \ln k'^2)$, or, less likely, as $e^{- k_T^2 / k^2 - k_T^2 / k'^2
}$. In any case, \eq{I1} suggests that each power of $k^2$ (or each
power of $k'^2$ or of $k \cdot k'$) gives a power of $1/\tau$ for
energy density $\epsilon$ in coordinate space: $k^2 \rightarrow
1/\tau$. (Indeed the powers of $k_T$ do not modify the
$\tau$-dependence of $T_{\mu\nu}$ at all.) Therefore, one may argue
that after the momentum integration is done in \eq{edens3}, the $f_1
(k^2, k'^2, k \cdot k', k_T) - f_1 (0,0,0, k_T)$ term, when
substituted into \eq{edens2}, yields approximately the following
contribution to $\epsilon$
\be\label{rem}
\frac{1}{\tau} \, \left[ f_1 \left(\frac{1}{\tau}, \frac{1}{\tau}, \frac{1}{\tau}, 
\langle k_T \rangle\right) - f_1 (0,0,0,\langle k_T \rangle) \right],
\ee
which falls off faster than $1/\tau$ and can thus be neglected
compared to \eq{ed}. This conclusion is natural, since the term in
\eq{rem}, or, equivalently, the second term on the right hand side 
of \eq{iter1}, does not contribute to the production cross section, as
follows from \eq{f6}, which is another way of saying that it is not
important at late times.

The proofs that the contributions to energy density $\epsilon$
generated by substituting $f_2 (k^2, k'^2, k \cdot k', k_T)$ $- f_2
(0,0,0,k_T)$ and $f_3 (k^2, k'^2, k \cdot k', k_T)$ $-f_3 (0,0,0,k_T)$
instead of $f_2$ and $f_3$ into \eq{edens2} are also subleading at
large $\tau$ can be constructed by analogy to the above.

Finally, we have to comment on our use of the Abelian part of
$T_{\mu\nu}$ only in \eq{tmnab} and throughout this Section. Including
the non-Abelian parts of the field strength tensor $F_{\mu\nu}$ would
generate higher powers of $A_\mu^a$ in the definition (\ref{tmnga}) of
$T_{\mu\nu}$. Using \eq{ffi} those extra powers can be rewritten as
extra integrals over $k''$ and $k'''$ in the extra terms which would
be added to \eq{tmnab}. Due to \eq{I1}, each of these extra integrals
would (at least) generate a Bessel function $J_{-\Delta} (k_T \,
\tau)$, which at large $\tau$ scales as $(1/\sqrt{\tau}) \, \cos (k_T
\, \tau + \frac{\pi}{2} \Delta - \frac{\pi}{4})$. Even without the cosine, one 
can immediately see that the cubic in $A_\mu^a$ term in $T_{\mu\nu}$
would fall off at least like $1/\tau^{3/2}$ at large $\tau$. The
quadric terms would fall off at least like $1/\tau^{2}$. Both of these
terms would be negligibly small compared to the leading quadratic term
scaling as $1/\tau$ shown in \eq{ed}.

\eq{ed} has a straightforward physical interpretation. Every Feynman 
diagram has a final state in which the particles are propagating as
non-interacting plane waves until the infinite late times. Indeed the
energy density of such a ``free-streaming'' state scales as $\sim
1/\tau$, and this is exactly what \eq{ed} represents.

Therefore, in this Section we have proven that in the
rapidity-independent case, defined by Eqs. (\ref{xmv3}) and
(\ref{xmv4}) for the total rapidity interval in the collision of two
very large nuclei, at any order in the perturbative expansion in the
strong coupling $g$, the resulting gluon field's energy density has a
non-vanishing term which is dominant at late times giving $\epsilon
\sim 1/\tau$ (\ref{ed}). Hence it appears that, in this boost-invariant case, 
thermalization leading to Bjorken hydrodynamic description of the
evolution of produced gluonic system, can not result from resummation
of perturbative QCD diagrams.

\section{Generalization to the Rapidity-Dependent Case}

For rapidity intervals $Y \, \gsim \, \frac{1}{\as}$ in heavy ion
collisions the quantum evolution corrections \cite{bfkl,dip,bk,jimwlk}
would become important making the produced particle distribution
rapidity dependent. Below we are first going to argue that
rapidity-dependent hydrodynamic description may only change the
$\epsilon \sim 1/\tau^{4/3}$ scaling of the ideal Bjorken energy
density to a higher power, $\epsilon \sim 1/\tau^{4/3 + \Delta}$. We
will then demonstrate that the rapidity-dependent quantum corrections,
such as the ones introduced by the BFKL evolution \cite{bfkl}, would
{\sl not} modify the $\epsilon \sim 1/\tau$ scaling derived in the
previous Section.

\subsection{Rapidity-Dependent Hydrodynamics}

In the rapidity dependent case the most general form of the
energy-momentum tensor is given by the equation similar to \eq{tmn1}
\be\label{tmnr1}
T_{\mu\nu} \, = \, A(\tau, \eta) \, u_\mu u_\nu + B(\tau, \eta) \,
(u_\mu v_\nu + u_\nu v_\mu) + C(\tau, \eta) \, v_\mu v_\nu + D(\tau, \eta)
\, g_{\mu\nu},
\ee
with $u_\mu$ and $v_\mu$ still given by Eqs. (\ref{umu}) and
(\ref{vmu}) and where now all the coefficients $A,B,C,D$ are also
functions of the space-time rapidity $\eta$. Due to this
$\eta$-dependence the $+ \leftrightarrow -$ symmetry argument no
longer applies in general. However, it still holds at mid-rapidity
($\eta = 0$) for a collision of two identical nuclei leading to
\be\label{Bcond}
B (\tau, \eta = 0) \, = \, 0.
\ee
Applying the conservation of energy-momentum tensor condition
(\ref{cons}) to the tensor in \eq{tmnr1} yields 
\begin{eqnarray}\label{eom1}
\tau \, \partial_\tau B - 2 \, \partial_\eta D + 2 \, \partial_\eta C + 2 \, B &=& 0 
\nonumber \\
2 \, \tau \, \partial_\tau A + \partial_\eta B + 2 \, \tau \,
\partial_\tau D + 2 \, A + 2 \, C &=& 0.
\end{eqnarray}
The energy-momentum tensor in \eq{tmnr1} would describe a hydrodynamic
system if it could be reduced to the standard hydrodynamic form
\be\label{hydro}
T_{\mu\nu} \, = \, (\epsilon + p) \, w_\mu \, w_\nu - p \, g_{\mu\nu},
\ee
where $w_\mu$ is the four-vector of the fluid velocity, $w_\mu \,
w^\mu = 1$. For the $1 + 1$-dimensional expansion of the system
created in a collisions of two very large nuclei considered here the
fluid velocity has zero transverse component, ${\un w} = 0$, such that
$w_\mu = (w_+, w_-, {\un 0})$. Matching \eq{tmnr1} onto \eq{hydro} we
obtain
\be\label{aep}
A \, = \, \epsilon + p + C
\ee
and
\be\label{dp}
D \, = \, - p.
\ee
For the hydrodynamic energy momentum tensor (\ref{hydro}) the
following relation holds
\be
T_{++} \, T_{--} \, = \, (T_{+-} + p)^2,
\ee
leading to a constraint 
\be\label{Cond}
C \, = \, \frac{B^2}{4 \, A}.
\ee
Combining \eq{eom1} and \eq{Cond} with the equation of state relating
$\epsilon$ and $p$, would give us a complete set of rapidity-dependent
hydrodynamic equations. However the resulting system of equations is
nonlinear and is hard to solve analytically. Instead we are going to
construct a perturbative solution for small rapidity-dependent
corrections to Bjorken hydrodynamics \cite{bj}. We begin by noting
that, since $B = 0$ in the boost-invariant case considered in the
previous Section, we can assume that non-zero $B$ reflects the
deviation from the ideal Bjorken hydrodynamics, and could be assumed
small if we are interested in small corrections to the
latter. Assuming that $B \ll A$ and keeping only linear in $B$
corrections allows us to neglect $C$, since, due to \eq{Cond}, $C \sim
B^2$. Than, using Eqs. (\ref{aep}) and (\ref{dp}) in \eq{eom1} yields
\begin{eqnarray}\label{eom2}
\tau \, \partial_\tau B + 2 \, \partial_\eta p + 2 \, B &=& 0 
\nonumber \\
2 \, ( \tau \, \partial_\tau \epsilon + \epsilon + p) + \partial_\eta
B \, &=& 0.
\end{eqnarray}
We are interested in the solution for the ideal gas equation of state:
therefore we put $\epsilon = 3 \, p$. The most general solution of
\eq{eom2} satisfying the condition of \eq{Bcond} and mapping back onto
Bjorken hydrodynamic behavior for small $B$ is
\be\label{esol}
\epsilon \, = \, \epsilon_0 \, \cos (\sqrt{\Delta} \, \eta) \, 
\frac{1}{\tau^{\frac{1}{3} \, (5 - \sqrt{1 - 3 \Delta})}}
\ee
with
\be\label{bsol}
B \, = \, - \frac{2}{3} \, \epsilon_0 \, \frac{\sqrt{1 - 3 \Delta} -
1}{\sqrt{\Delta}} \, \sin (\sqrt{\Delta} \, \eta) \,
\frac{1}{\tau^{\frac{1}{3} \, (5 - \sqrt{1 - 3 \Delta})}},
\ee
where $\Delta$ and $\epsilon_0$ are arbitrary constants. The
corresponding flow velocity components are given by
\be\label{wpm}
w_\pm \, \approx \, \frac{x_\pm}{\tau} \, \left( 1 \pm \frac{3 \, B}{8 \,
\epsilon} \right).
\ee
Looking at the solution given by \eq{esol} one may wonder why the
energy density is not positive definite. Indeed for $\Delta < 0$ the
energy density $\epsilon$ from \eq{esol} becomes positive definite,
since $\cos (\sqrt{\Delta} \, \eta)$ would be replaced by $\cosh
(\sqrt{|\Delta|} \, \eta)$. However, the resulting rapidity
distribution of energy density would increase as one moves further
away from mid-rapidity, which is unphysical. Therefore one has to have
$\Delta > 0$. Resolution of the positivity problem for $\epsilon$
comes from the necessity to satisfy the $B \ll A$ assumption which we
have made at the beginning of this calculation. It translates into $B
\ll \epsilon$ condition, which is satisfied by Eqs. (\ref{esol}) and 
(\ref{bsol}) only if $\sqrt{\Delta} \, \eta \ll 1$. Since in this
Section we are interested in large rapidity intervals, $\eta \sim Y
\gsim 1/\as$, the $\sqrt{\Delta} \, \eta \ll 1$ requires that 
$\Delta \, \lsim \, \as \ll 1$. Hence, for large rapidities,
Eqs. (\ref{esol}) and (\ref{bsol}) are valid only at the lowest order
in $\Delta$
\be\label{esole}
\epsilon \, \approx \, \frac{\epsilon_0}{\tau^{\frac{4}{3} + \frac{\Delta}{2}}} \, 
\left(1 - \frac{1}{2} \, \Delta \, \eta^2 \right),
\ee
\be\label{bsole}
B \, \approx \, \frac{\epsilon_0}{\tau^{\frac{4}{3} +
\frac{\Delta}{2}}} \, \Delta \, \eta,
\ee
where we did not expand $\tau^{-\Delta/2}$ since, at late times, $\Delta
\, \ln \tau$ does not have to be small for $B \ll \epsilon$ condition 
to hold. For small $\sqrt{\Delta} \, \eta$ the energy density in
\eq{esole} is indeed positive.

\eq{esole} has an important feature which we would like to emphasize: 
since $\Delta > 0$, it shows that the energy density of the
boost-non-invariant ideal hydrodynamics falls off with $\tau$ {\sl
faster} than the energy density of the boost-invariant ideal Bjorken
hydrodynamics \cite{bj}. Here we have proven it only for a small
rapidity-dependent perturbation of the Bjorken solution. However one
should expect our conclusion to hold in a general case of a
rapidity-dependent hydrodynamics. In the case of a rapidity-dependent
hydrodynamics, the longitudinal pressure is higher than in the
boost-invariant Bjorken case, generating the longitudinal acceleration
of the flow (see e.g. \eq{wpm}). The central-rapidity high-density
system starts expanding faster than in Bjorken case, leading to a
faster depletion of the energy density with $\tau$. In other words,
once the longitudinal homogeneity of pure Bjorken hydrodynamics is
broken by some rapidity-dependent phenomena, the system starts doing
more work in the longitudinal direction than it was doing in pure
Bjorken hydrodynamics case, and this leads to a faster decrease of
energy density with proper time.\footnote{The author would like to
thank Ulrich Heinz for explaining to him this argument.}

\subsection{Rapidity-Dependent Energy Density}

\label{arg2}

Here we are going to generalize the argument of Sect. \ref{arg1} to
the case of rapidity-dependent gluon fields. It is impossible to
define co-moving energy density for a general energy-momentum tensor
like the one given in \eq{tmnr1}, since, in the general not
necessarily hydrodynamic case, one can not define the co-moving frame,
and in the case of hydrodynamics (\ref{hydro}) one needs to know the
flow velocity to define the co-moving frame, which is impossible to do
without solving the hydrodynamics equations (\ref{eom1}). Therefore we
will restrict our analysis to the case of mid-rapidity, $\eta = 0$,
where, for a collision of two identical nuclei, the co-moving frame is
just the center of mass frame of the two nuclei. There \eq{edens}
would apply, such that
\be\label{edense}
\epsilon (\tau , \eta=0) \, = \, \frac{1}{2} \, \left( \frac{\tau}{x_+} \right)^2 \, 
T_{++} (\tau , \eta=0) + T_{+-} (\tau , \eta=0) \, = \, T_{++} (\tau ,
\eta=0) + T_{+-} (\tau , \eta=0).
\ee
Repeating the steps from Section \ref{arg1} we write
\ben
\epsilon (\tau , \eta=0) \, = \, \int \frac{ d^4 k \, d^4 k'}{(2 \pi)^8}\,
\frac{e^{-i k \cdot x - i k' \cdot x}}{(k^2 + i \epsilon k_0) \, (k'^2
+ i \epsilon k'_0)}\bigg|_{\eta = 0} \, \left\{ \left[ \frac{1}{2} \, \left(
\frac{\tau}{x_+} \right)^2 \, k_+ \, k'_+ - \frac{1}{2} \, {\un k}^2
\right] \,  \right.
\een
\ben
\times \, f_1 (k^2, k_+, k'^2, k'_+, k \cdot k', k_T) \, +  \left[ 2 \, k_- \, k'_- - \frac{1}{2} \, \left(
\frac{\tau}{x_+} \right)^2 \, {\un k}^2 \right] \, k_+ \, k'_+ \, f_2 (k^2, k_+, k'^2, k'_+, k \cdot k', k_T) 
\een
\be\label{edense1}
+ \left. \left[ - \frac{1}{4} \, \left( \frac{\tau}{x_+} \right)^2 \,
\left( \frac{k'_+}{k_-} + \frac{k_+}{k'_-} \right) \, + \, 1 \right] 
\, f_3 (k^2, k_+, k'^2, k'_+, k \cdot k', k_T) \right\} \, 
\frac{(2 \pi)^2}{S_\perp} \, \delta ({\un k} + {\un k}').
\ee
where now, in the rapidity dependent case, $f_i$'s are functions of
$k_\pm$ and $k'_\pm$ as well. However, since we can always rewrite
$k_- = (k^2 + k_T^2)/2 k_+$ and $k'_- = (k'^2 + k_T^2)/2 k'_+$, we put
only $k_+$ and $k'_+$ in the arguments of the functions $f_i$.

Rapidity-dependent quantum evolution corrections come in as logarithms
of Bjorken $x$ variable \cite{bfkl}. If $p_+$ is a large longitudinal
momentum carried by a nucleon in the nucleus moving in the
$+$-direction, than $x = k_+ / p_+$. The rapidity-dependent
corrections would then bring in powers of $\as \, \ln 1/x \, = \, \as
\, \ln p_+ / k_+$. Resummation of such corrections for the gluon production 
amplitudes generates powers of $1/x$, or, equivalently, $p_+ /
k_+$. Therefore, to verify whether such corrections modify the
$\tau$-dependence of $\epsilon$, we can consider the following general
form for the functions $f_i$'s
\be\label{frap}
f_i (k^2, k_+, k'^2, k'_+, k \cdot k', k_T) \, = \, \left( \frac{p_+}{k_+}
\, \frac{p_+}{k'_+} \right)^{\lambda} \, 
\tilde{f}_i (k^2, k'^2, k \cdot k', k_T),
\ee
where we again used the fact that $f$'s are symmetric under $k
\leftrightarrow k'$ interchange. For simplicity we assume the 
power $\lambda$ to be the same for $f_1$, $f_2$ and $f_3$: this
assumption is not crucial and can be easily relaxed. The logarithmic
corrections to $f_i$'s, i.e., terms with $\ln p_+ / k_+$ and $\ln p_+
/ k'_+$, can be obtained from $f_i$'s in \eq{frap} by differentiating
it with respect to $\lambda$. Indeed that would give only logarithms
like $\ln (p_+^2/k_+ k'_+)$, but not $\ln k_+/k'_+$: while we are
quite confident that the latter terms never appear in perturbation
theory, our approach can be easily generalized to include both types
of logarithms by putting different powers for $p_+/k_+$ and $p_+/k'_+$
factors in \eq{frap}. (A careful reader may worry that the
$p_+$-dependence was never explicitly shown in the argument of $f_i$'s
and was explicitly assumed there: in fact, due to boost-invariance,
the $p_+$-dependence enters in $f_i$'s only through the ratios of $p_+
/ k_+$ and $p_+ / k'_+$. In the rapidity-independent case of Section
\ref{arg1}, $f_i$'s were independent of $p_+$, which corresponds to the
eikonal limit.)

Substituting $f_i$'s from \eq{frap} into \eq{edense1} we obtain
\ben
\epsilon (\tau , \eta=0) \, = \, \int \frac{ d^4 k \, d^4 k'}{(2 \pi)^8}\,
\frac{e^{-i k \cdot x - i k' \cdot x}}{(k^2 + i \epsilon k_0) \, (k'^2
+ i \epsilon k'_0)}\bigg|_{\eta = 0} \, \left\{ \left[ \frac{1}{2} \, \left(
\frac{\tau}{x_+} \right)^2 \, k_+ \, k'_+ - \frac{1}{2} \, {\un k}^2
\right] \,  \right.
\een
\ben
\times \,  
\tilde{f}_1 (k^2, k'^2, k \cdot k', k_T) \, +  \left[ 2 \, k_- \, k'_- - \frac{1}{2} \, \left(
\frac{\tau}{x_+} \right)^2 \, {\un k}^2 \right] \, k_+ \, k'_+ \, \
\tilde{f}_2 (k^2, k'^2, k \cdot k', k_T) +
\een
\be\label{edense2}
+ \left. \left[ - \frac{1}{4} \, \left( \frac{\tau}{x_+} \right)^2 \,
\left( \frac{k'_+}{k_-} + \frac{k_+}{k'_-} \right) \, + \, 1 \right] 
\, \tilde{f}_3 (k^2, k'^2, k \cdot k', k_T) \right\} \, \left( \frac{p_+}{k_+}
\, \frac{p_+}{k'_+} \right)^{\lambda} \, \frac{(2 \pi)^2}{S_\perp} \, \delta ({\un k} + {\un k}').
\ee
To perform the longitudinal momentum integrals we will use the
integral in Eqs. (\ref{J1}) and (\ref{J6}) of Appendix B. There we can
see that, similar to the rapidity-independent case, each positive
extra power of $k^2$ (or $k'^2$ or $k \cdot k'$) gives a power of
$1/\tau$. Therefore, we again are interested in contribution of $f_i
(k^2 =0, k_+, k'^2 =0, k'_+, k \cdot k'=0, k_T)$ as in the terms
giving the leading-$\tau$ behavior. Similar to \eq{f7} we can write
\be\label{fe7}
\frac{1}{S_\perp} \, \left( \frac{p_+}{k_+}
\, \frac{p_+}{- k_+} \right)^{\lambda} \, \left[ \tilde{f}_1 (0, 0, 0, k_T) 
- k_T^2 \tilde{f}_2 (0, 0, 0, k_T) -
\frac{2}{k_T^2} \, \tilde{f}_3 (0, 0, 0, k_T) \right] \,
= \, - 2 (2 \pi)^3 \, 
\frac{dN}{d^2 k \, dy \, d^2 b}.
\ee
which shows that this combination of $f_i$'s is not zero.  Using
\eq{fe7} in \eq{edense2}, where we put $k^2 = k'^2 =0$ in the 
arguments of all $f_i$'s, and performing the longitudinal integrations
using the formulas from Appendix B yields for the leading term in
energy density
\be\label{edense3}
\epsilon (\tau , \eta=0) \, \approx \, \frac{\pi}{2} \,  \int d^2 k \, 
\frac{d N}{d^2 k \, d \eta \, d^2 b}\bigg|_{\eta = 0} \ k_T^2 \, 
\left\{ \left[ J_{-1-\lambda} (k_T \tau) \right]^2 + \left[ J_{-\lambda} 
(k_T \tau) \right]^2 \right\}.
\ee
In arriving at \eq{edense3} we have noticed that, according to
\eq{J6}, each power of $k_+$ or $k'_+$ gives a power of $k_T \, e^\eta /\sqrt{2}$ 
after the integration. For on-mass shell gluons in \eq{fe7} one has
$k_+ = k_T \, e^y /\sqrt{2}$. Therefore, the powers of $k_T \, e^\eta
/\sqrt{2}$ were absorbed in $\frac{d N}{d^2 k \, d \eta \, d^2 b}$ by
just replacing $y \rightarrow \eta$.

The large-$\tau$ asymptotics of \eq{edense3} is the same as in
\eq{ed}:
\be\label{ede}
\epsilon (\tau , \eta=0) \bigg|_{\tau \gg 1/\langle k_T \rangle} \, \approx \, 
\frac{1}{\tau} \, \int d^2 k \,  \frac{d N}{d^2 k \, d \eta \, d^2 b}\bigg|_{\eta = 0}
 \ k_T \, = \, \frac{1}{\tau} \, \frac{d E_T}{d \eta \, d^2
 b}\bigg|_{\eta = 0}.
\ee
Therefore, we have proven that even in the rapidity-dependent case the
mid-rapidity energy density given by the Feynman diagrams falls off as
$1/\tau$ at large $\tau$. This conclusion could be easily derived by
just analyzing Eqs. (\ref{J1}) and (\ref{J6}): one can see there that
each extra power of $k_+$ does not bring in any new powers of $\tau$,
and only modifies the order of the Bessel function, which can not
change the $\tau$-dependence, as follows from \eq{edense3}.

\eq{ede} shows that the scaling of energy density of the 
rapidity-dependent solution of the hydrodynamics equations given by
\eq{esole}, $\epsilon \sim 1/\tau^{\frac{4}{3} + \frac{\Delta}{2}}$,  
can not come from the leading contribution of Feynman
diagrams. Indeed, the subleading contributions may still lead to
energy density falling off with $\tau$ faster than $1/\tau$: however,
due to \eq{I1}, such contributions must come in with extra positive
powers of $k^2$ in momentum space. They would go to zero in the
on-mass shell $k^2 \rightarrow 0$ limit and, thus, would not
contribute to the production cross section. Such corrections are
probably irrelevant for all physical observables.

\section{Including Quarks}

\label{arg3}

To generalize our conclusion to massless quarks we will restrict our
discussion to rapidity-independent case only: generalization to
the rapidity-dependent case can be easily done following the procedure
outlined in Section \ref{arg2}. We start with the energy-momentum
tensor for a single massless quark flavor:
\be\label{tmnq}
T_{\mu\nu}^{quark} \, = \, \frac{i}{2} \, \overline{\psi} \,
(\gamma_\mu \, D_\nu + \gamma_\nu \, D_\mu) \, \psi.
\ee
The corresponding energy density is given by \eq{edens}, which we
again want to rewrite as a double integral, just like \eq{edens2}, by
Fourier transforming the quark field
\be\label{psi1}
\psi (x) \, = \, i \, \int \frac{d^4 k}{(2 \pi)^4} \, 
\frac{k \cdot \gamma}{k^2 + i \epsilon k_0} \, e^{-i k \cdot x} \, \xi (k)
\ee
and
\be\label{psi2}
\overline{\psi} (x) \, = \, i \, \int \frac{d^4 k'}{(2 \pi)^4} \, 
\tilde{\xi} (k') \, \frac{k' \cdot \gamma}{k'^2 + i \epsilon k'_0} \, e^{-i k' \cdot x} 
\ee
with $\xi (k)$ and $\tilde{\xi} (k')$ some spinors.  In the following,
similar to Section \ref{arg1}, we will keep only the Abelian part of
$T_{\mu\nu}^{quark}$. The non-Abelian corrections are suppressed at
late times and can be neglected, since, just like in Section
\ref{arg1}, they fall off faster than the Abelian term by at least a
factor of $1/\sqrt{\tau}$. Replacing the covariant derivatives $D_\mu$
in \eq{tmnq} by a regular derivative $\partial_\mu$ and substituting
Eqs. (\ref{psi1}) and (\ref{psi2}) in it we obtain
\be\label{tmnq1}
T_{\mu\nu}^{quark} \, = \, - \frac{1}{2} \, \int \frac{ d^4 k \, d^4
k'}{(2 \pi)^8} \, \frac{e^{-i k \cdot x - i k' \cdot x}}{(k^2 + i
\epsilon k_0) \, (k'^2 + i \epsilon k'_0)} \, \left< \tilde{\xi} (k') \, 
k' \cdot \gamma \, (\gamma_\mu \, k_\nu + \gamma_\nu \, k_\mu) \, k
\cdot \gamma \, \xi (k) \right>.
\ee
Rewriting
\be\label{xhh}
\left<\left< \tilde{\xi} (k') \, k' \cdot \gamma \, \gamma_\mu \, k
\cdot \gamma \, \xi (k) \right>\right> =  - \left[ (k_\mu - k'_\mu) \, h_1 (k^2,
k'^2, k \cdot k', k_T) + (k_\mu + k'_\mu) \, h_2 (k^2, k'^2, k \cdot k', k_T) \right]
\ee
in \eq{tmnq1} and using \eq{edens} we write
\ben
\epsilon^{quark} (\tau) \, = \, 
\int \frac{ d^4 k \, d^4 k'}{(2 \pi)^8}\,
\frac{e^{-i k \cdot x - i k' \cdot x}}{(k^2 + i \epsilon k_0) \, (k'^2
+ i \epsilon k'_0)} \, 
\een
\ben
\times \, \Bigg\{ \Bigg[ \frac{1}{2} \, \left(
\frac{\tau}{x_+}\right)^2 \, (k_+ - k'_+) \, k_+ + k_+ \, k_- 
- \frac{1}{2} \, (k'_+ \, k_- + k_+ \, k'_-) \Bigg] \, h_1 (k^2,
k'^2, k \cdot k', k_T) \, + 
\een
\be\label{qedens1}
+ \, \Bigg[ \frac{1}{2} \, \left(
\frac{\tau}{x_+}\right)^2 \, (k_+ + k'_+) \, k_+ + k_+ \, k_- + \frac{1}{2} \, 
(k'_+ \, k_- + k_+ \, k'_-) \Bigg] 
\, h_2 (k^2, k'^2, k \cdot k', k_T) \, \Bigg\} \, \frac{(2 \pi)^2}{S_\perp} \, \delta ({\un k}
+ {\un k}').
\ee
Similar to Section \ref{arg1}, by putting $k = - k'$ and $k^2 = k'^2
=0$ in \eq{xhh} we derive
\be\label{h1}
\frac{1}{S_\perp} \, h_1 (k^2 =0, k'^2 =0, k \cdot k' =0, k_T) \, = \, \frac{1}{S_\perp} 
\left<\left< \tilde{\xi} (-k) \,
k \cdot \gamma \, \xi (k) \right>\right>\bigg|_{k^2 =0} \, = \, 2 (2
\pi)^3 \,
\frac{dN^q}{d^2k \, dy \, d^2 b}, 
\ee
where $\frac{dN^q}{d^2k \, dy \, d^2 b}$ is the multiplicity of the
produced quarks.  Arguing, just like we did for gluons, that the
leading-$\tau$ behavior for the quark energy density is given by $h_1
(k^2 =0, k'^2 =0, k \cdot k' =0, k_T)$ and $h_2 (k^2 =0, k'^2 =0, k
\cdot k' =0, k_T)$ in \eq{qedens1} we can integrate over the longitudinal 
momenta in \eq{qedens1} using \eq{J7} in Appendix B obtaining
\ben
\epsilon^{quark} (\tau ) \, \approx \, \frac{\pi}{2} \, \int d^2 k \, 
\frac{dN^q}{d^2k \, d\eta \, d^2 b} \, k_T^2 \, \left\{ - J_0 (k_T \tau) \, 
J_2 (k_T \tau) + 2 [J_1 (k_T \tau)]^2 + [J_0 (k_T \tau)]^2 \right\} + 
\een
\be\label{qedens2}
+ \, \frac{1}{8 \, S_\perp} \, \int \frac{d^2 k}{(2 \, \pi)^2} \, h_2
(0, 0, 0, k_T) \, k_T^2 \, 
\left\{ - J_0 (k_T \tau) \, J_2
(k_T \tau) - 2 [J_1 (k_T \tau)]^2 + [J_0 (k_T \tau)]^2 \right\},
\ee
where we also substituted space-time rapidity $\eta$ instead of $y$ in
the quark multiplicity distribution, which makes no difference in the
boost invariant case we consider.  We can assume that $h_2 (k^2 =0,
k'^2 =0, k \cdot k' =0, k_T)$ is a steeply falling function of $k_T$ for $k_T \gg
\langle k_T \rangle \sim Q_s$. The assumption is justified since $h_2 (k^2 =0,
k'^2 =0, k \cdot k' =0, k_T)$ comes from the same amplitude that gave $h_1 (k^2 =0,
k'^2 =0, k \cdot k' =0, k_T)$, which is equal to the quark spectrum, as shown in
\eq{h1}, which in turn is always a steeply falling function of
$k_T$ scaling at least like $\sim 1/k_T^4$ for $k_T$ above some scale
$\langle k_T \rangle \sim Q_s$. By the same argument $h_2 (k^2 =0,
k'^2 =0, k \cdot k' =0, k_T)$ should be regular (or at most
logarithmically divergent) at $k_T = 0$.  Using these assumptions we
can argue that the integral in the second term on the right hand side
of \eq{qedens2} is dominated by $k_T \sim Q_s$, which allows us to
rewrite it as
\be\label{t2}
\frac{1}{16 \, \pi \, S_\perp} \,  h_2 (0, 0, 0, Q_s) \, 
\int_0^{Q_s} dk_T \, k_T^3 \, \left\{ - J_0 (k_T \tau) \, J_2
(k_T \tau) - 2 [J_1 (k_T \tau)]^2 + [J_0 (k_T \tau)]^2 \right\}.
\ee
Performing the integration in \eq{t2} one would obtain a linear
combination of hypergeometric functions, which can be shown to fall
off at least as $\sim 1/\tau^2$ at large $\tau$. Therefore the second
term on the right hand side of \eq{qedens2} falls off with $\tau$ at
least as $\sim 1/\tau^2$, and can be neglected if the first term on
the right hand side of \eq{qedens2} falls off with $\tau$ slower than
$\sim 1/\tau^2$.

This can be easily verified. At large $\tau$ the first term on the
right hand side of \eq{qedens2} gives
\be\label{qed}
\epsilon^{quark} (\tau )\bigg|_{\tau \gg 1/\langle k_T \rangle} \, \approx \, 
\frac{2}{\tau} \, \int d^2 k \, \frac{dN^q}{d^2k \, d\eta \, d^2 b} \, k_T \, = \, 
\frac{1}{\tau} \, \frac{dE_T^{quarks}}{d\eta \, d^2 b}, 
\ee
since the factor of $2$ accounts for the anti-quark contribution. We
can see that indeed the term in \eq{t2} is negligibly small compared
to \eq{qed}, which gives us the dominant contribution to the energy
density due to quarks at late times $\tau$. (Of course the typical
transverse momentum $\langle k_T \rangle$ in \eq{qed} does not have to be exactly
equal to the similar typical momentum for gluons in \eq{ed}: however,
the difference between the two is usually given by the ratio of the
Casimir operators, which is just a constant ($4/9$) and does not
change our argument above.)

We have shown that inclusion of massless quarks does not change the
conclusion of the previous Sections that the leading diagrammatic
contribution to energy density scales as $1/\tau$ at large $\tau$ for
any order in the coupling $g$. Therefore, it appears that inclusion of
quarks does not affect the onset of thermalization.

\section{Conclusions}

We have shown above in Eqs. (\ref{ed}) and (\ref{qed}) that gluon and
quark fields generated by Feynman diagrams in high energy heavy ion
collisions lead to energy densities scaling as $\epsilon \sim 1/\tau$
at $\tau~\gg~1/\langle k_T \rangle$.  In the saturation/Color Glass picture of heavy
ion collisions, the typical momentum $\langle k_T \rangle$ is proportional to the
saturation scale $Q_s$. Therefore, the $1/\tau$ scaling of energy
density sets in at relatively {\sl early} times $\tau \sim 1/Q_s$
(even though throughout the paper we called these proper times {\sl
late} times). The remaining evolution of the system in the
perturbative scenario considered above, characterized by $\epsilon
\sim 1/\tau$ scaling, is reminiscent of the so-called {\sl free streaming}, 
where the system simply falls apart without particles interacting. 

%%%%%%%%%%%%%%%%%%%%%%%%%%%%%
\begin{figure}[b]
\begin{center}
\epsfxsize=15cm
\leavevmode
\hbox{\epsffile{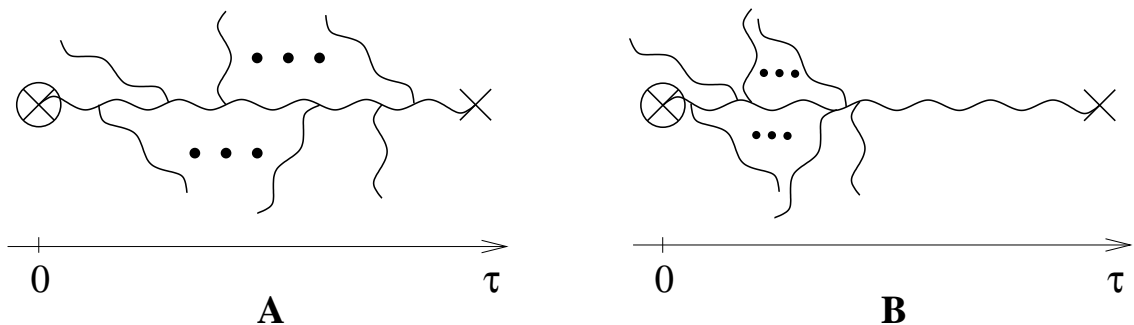}}
\end{center}
\caption{(A) Gluon field produced in a collision with all the interactions 
throughout its proper time evolution. (B) The dominant contribution to
the gluon field comes from early-time interactions. }
\label{frst}
\end{figure}
%%%%%%%%%%%%%%%%%%%%%%%%%%%%%%

To explain how this happens, let us provide a diagrammatic
interpretation of our conclusion of $\epsilon \sim 1/\tau$
scaling. Let us imagine a general gluon field produced in a heavy ion
collision as shown in \fig{frst}A. There the gluon field is first
produced in the nuclear collision at $\tau = 0$ denoted by the
$\otimes$ sign. The cross at the other end denotes the later point in
$\tau$ where we measure the energy density of the gluon field. In the
evolution of the system the gluon interacts with other gluon fields
produced in the collision in all possible ways, as shown in
\fig{frst}A. However, the proper times of these interactions are not fixed: 
they are integrated over the whole range of $\tau$. Interactions may
also happen at different impact parameters, which are also integrated
over. The $\epsilon \sim 1/\tau$ scaling conclusion from
Eqs. (\ref{ed}) and (\ref{qed}) appears to indicate that the dominant
diagrams are given by \fig{frst}B, where all the interactions happen
at early times, after which the system simply falls apart. In other
words, the integrations over proper times of the interactions in
\fig{frst}A are dominated by early times of \fig{frst}B. $\langle k_T \rangle$, or
$Q_s$, being the only scale in the problem, sets the typical time
scale for the end of interaction period and the onset of free
streaming, $\tau_0 \sim 1/Q_s$. Such behavior has been previously
observed in the numerical simulations of the classical gluon fields
\cite{KV,KNV}.

Another way to physically understand our conclusion of $\epsilon \sim
1/\tau$ scaling is as follows.  At any order in the coupling constant
$\as$ the gluon (or quark) field has a diagram (or several diagrams)
which is (are) non-zero if the gluon is put on mass-shell. This is just a statement
that gluon (or quark) multiplicity distribution can be expanded in a
perturbation series in $\as$. As we have shown above in Sections
\ref{arg1}, \ref{arg2} and \ref{arg3}, such diagrams always give energy 
density scaling as $1/\tau$. Each diagram is dominated by the on-shell
particles free streaming away, which always leads to $\epsilon \sim
1/\tau$.

Therefore we have shown that the onset of thermalization and the
subsequent Bjorken or rapidity-dependent hydrodynamic expansion of the
system of quarks and gluons produced in heavy ion collisions can not
result from summation of Feynman diagrams. Nevertheless, there
exists a solid phenomenological evidence for the strong final state
interactions
\cite{dAtaphen,dAtaphob,dAtastar,brahms,aaphenix,aaphobos,aastar,Bj,EL,BDMPSfull,EL2,Zak,SW}
and hydrodynamic behavior
\cite{EKR,hydro1,hydro2,HN} of the system produced in heavy ion collisions 
at RHIC, indicating a formation of strongly interacting quark-gluon
plasma (QGP). To reconcile it with the above argument that Feynman
diagrams lead to a free streaming behavior for both quarks and gluons,
one must conclude that non-perturbative QCD effects are instrumental
in QGP formation at RHIC. These could be the the non-perturbative
effects associated with infrared modes having momenta of the order of
$\Lambda_{QCD}$ which can not be represented by Feynman
diagrams. Therefore, the above argument does not apply to such
modes. Alternatively, the non-perturbative effects might be of the
nature similar to the ultra-soft modes in finite temperature
non-Abelian field theories, which have momenta of the order of $g^2 \,
T$ with $T$ the temperature of the system. It is well-known that
resummation of ultra-soft modes is a non-perturbative problem in
finite temperature QCD \cite{Linde}. It is also known that ultra-soft
modes are very important for many physical observables for equilibrium
QCD matter at finite temperature \cite{Bodeker,ASY,BI}. If they are
important for equilibrium QCD matter, it would be natural to suggest
that the ultra-soft modes could also play a major role in
non-equilibrium phenomena such as the onset of
thermalization. However, a more careful analysis of the issue is
needed in order to draw any conclusions. Such analysis is beyond the
scope of this paper.

Distinguishing which one of the two types of non-perturbative effects
plays a more important role in the process of thermalization would
also be important for our understanding of LHC heavy ion data. The
non-perturbative effects characterized by the scale $\Lambda_{QCD}$
are likely to be of little importance at LHC where the saturation
scale $Q_s$ is predicted to be much larger than $\Lambda_{QCD}$
shifting most partons away from the infrared region. At the same time,
the non-perturbative ultra-soft modes carrying momenta $g^2 \, T$ may
remain important even at high LHC energies if the relevant temperature
scales with the saturation scale, $T \sim Q_s$, increasing at high
energy.

\section*{Acknowledgments}

The author would like to thank Ian Balitsky, Eric Braaten, Ulrich
Heinz, Larry McLerran, Al Mueller and Dam Son for many informative
discussions. The author is also grateful to Ulrich Heinz for
proofreading the manuscript.

This work is supported in part by the U.S. Department of Energy under
Grant No. DE-FG02-05ER41377.

\renewcommand{\theequation}{A\arabic{equation}}
  \setcounter{equation}{0} 
  \section*{Appendix A}

Here we are going to prove the following formula
\be\label{I1}
I \, \equiv \, \int_{-\infty}^{\infty} d k_+ \, dk_- \, e^{- i k_+ x_-
- i k_- x_+} \, (k^2 + i \epsilon k_0)^{\Delta - 1} \, = \, - \frac{2
\pi^2}{\Gamma (1-\Delta)} \, \left( \frac{2 \, k_T}{\tau} \right)^\Delta
\, e^{i \, \pi \, \Delta} \, J_{-\Delta} (k_T \tau)
\ee
with $k^2 = 2 \, k_+ \, k_- - {\un k}^2$ and for $x_+ > 0$, $x_- > 0$,
and $\Delta >0$. Let us first rewrite the integral (\ref{I1}) as
\be\label{I2}
I \, = \, 2^{\Delta - 1} \, \int_{-\infty}^{\infty} \frac{d k_+ \,
dk_- \, e^{- i k_+ x_- - i k_- x_+}}{(k_+ + i \epsilon )^{1-\Delta}} \,
\left( k_- - \frac{{\un k}^2}{2 (k_+ + i \epsilon)} + i \epsilon
\right)^{\Delta - 1}.
\ee
Defining $\tilde{k}_- = k_- - {\un k}^2 / 2 k_+$ we write
\be\label{I3}
I \, = \, 2^{\Delta - 1} \, \int_{-\infty}^{\infty} \frac{d k_+}{(k_+
+ i \epsilon )^{1-\Delta}} \, e^{- i k_+ x_- - i \frac{{\un k}^2}{2
k_+ + i \epsilon} \, x_+} \, \int_{-\infty}^{\infty} d\tilde{k}_- \,
e^{- i \tilde{k}_- \, x_+} \, ( \tilde{k}_- + i \epsilon )^{\Delta - 1}.
\ee
The $\tilde{k}_-$ integral can be easily performed by distorting the
integration contour around the branch cut. We obtain
\be\label{I4}
I \, = \, -  2^{\Delta - 1} \, \frac{2 \pi i \, e^{i \frac{\pi}{2}
\Delta}}{\Gamma (1-\Delta)} \, x_+^{-\Delta} \, \int_{-\infty}^{\infty} \frac{d
k_+}{(k_+ + i \epsilon )^{1-\Delta}} \, e^{- i k_+ x_- - i \frac{{\un
k}^2}{2 k_+ + i \epsilon} \, x_+}.
\ee
Expanding the second term in the power of the exponent in \eq{I4} in a
Taylor series we write
\be\label{I5}
I \, = \, -  2^{\Delta - 1} \, \frac{2 \pi i \, e^{i \frac{\pi}{2}
\Delta}}{\Gamma (1-\Delta)} \, x_+^{-\Delta} \, \sum_{n=0}^\infty \, \frac{1}{n!} \, 
\left(\frac{- i {\un k}^2 \, x_+}{2}\right)^n \int_{-\infty}^{\infty} 
\frac{d k_+}{(k_+ + i \epsilon )^{n + 1-\Delta}} \, e^{- i k_+ x_-}.
\ee
Performing the $k_+$ integration just like we did the $\tilde{k}_-$
integral above yields
\be\label{I6}
I \, = \, - 2^{\Delta - 1} \, \frac{(2 \pi)^2  \, e^{i \pi
\Delta}}{\Gamma (1-\Delta)} \, (x_+ \, x_-)^{-\Delta} \, \sum_{n=0}^\infty \, 
\frac{1}{n! \, \Gamma (n+1-\Delta)} \, 
\left(\frac{- {\un k}^2 \, x_+ \, x_-}{2}\right)^n.
\ee
Remembering that $2 x_+ x_- = \tau^2$ and performing the summation
over $n$ we obtain \eq{I1} as desired.

\renewcommand{\theequation}{B\arabic{equation}}
  \setcounter{equation}{0} \section*{Appendix B}

Our goal in this appendix is to perform the following integration
\be\label{J1}
J \, \equiv \, \int_{-\infty}^{\infty} d k_+ \, dk_- \, e^{- i k_+ x_-
- i k_- x_+} \, (k^2 + i \epsilon k_0)^{\Delta - 1} \, (k_+ + i
\epsilon)^\lambda. 
\ee
Repeating the steps from Appendix A which led to \eq{I4} we write
\be\label{J2}
J \, = \, -  2^{\Delta - 1} \, \frac{2 \pi i \, e^{i \frac{\pi}{2}
\Delta}}{\Gamma (1-\Delta)} \, x_+^{-\Delta} \, \int_{-\infty}^{\infty} \frac{d
k_+}{(k_+ + i \epsilon )^{1-\Delta - \lambda}} \, e^{- i k_+ x_- - i \frac{{\un
k}^2}{2 k_+ + i \epsilon} \, x_+}.
\ee
Expanding the second term in the exponent yields
\be\label{J3}
J \, = \, -  2^{\Delta - 1} \, \frac{2 \pi i \, e^{i \frac{\pi}{2}
\Delta}}{\Gamma (1-\Delta)} \, x_+^{-\Delta} \, \sum_{n=0}^\infty \, \frac{1}{n!} \, 
\left(\frac{- i {\un k}^2 \, x_+}{2}\right)^n \int_{-\infty}^{\infty} 
\frac{d k_+}{(k_+ + i \epsilon )^{n + 1-\Delta-\lambda}} \, e^{- i k_+ x_-}.
\ee
Performing the $k_+$-integration we obtain
\be\label{J4}
J \, = \, - 2^{\Delta - 1} \, \frac{(2 \pi)^2  \, e^{i \pi
\Delta + i \frac{\pi}{2} \lambda}}{\Gamma (1-\Delta)} \, 
(x_+ \, x_-)^{-\Delta} \, x_-^{-\lambda} \, \sum_{n=0}^\infty \, 
\frac{1}{n! \, \Gamma (n+1-\Delta-\lambda)} \, 
\left(\frac{- {\un k}^2 \, x_+ \, x_-}{2}\right)^n,
\ee
which, after summing over $n$ gives
\be\label{J5}
J \, = \,  - \frac{2 \pi^2}{\Gamma (1-\Delta)} \, 
\left( \frac{2 \, k_T}{\tau} \right)^\Delta
\, \left( \frac{k_T \, \tau}{2 \, x_-} \right)^\lambda
\, e^{i \, \pi \, \Delta + i \frac{\pi}{2} \lambda} \, 
J_{-\Delta - \lambda} (k_T \tau).
\ee
Noting that $x_\pm = \tau e^{\pm \eta} /\sqrt{2}$ with $\eta$ the
space-time rapidity we rewrite \eq{J5} as
\be\label{J6}
J \, = \,  - \frac{2 \pi^2}{\Gamma (1-\Delta)} \, 
\left( \frac{2 \, k_T}{\tau} \right)^\Delta
\, \left( \frac{k_T}{\sqrt{2}} \right)^\lambda \, e^{\lambda \, \eta}
\, e^{i \, \pi \, \Delta + i \frac{\pi}{2} \lambda} \, J_{-\Delta - \lambda} (k_T \tau).
\ee
As one can see from \eq{J6}, extra powers of $k_+$ in \eq{J1} as
opposed to \eq{I1} do not bring in any extra inverse powers of $\tau$:
they only modify the order of the Bessel function.

Finally, let us list here another useful integral, which can be easily
obtained by direct integration
\be\label{J7}
\int_{-\infty}^{\infty} d k_+ \, dk_- \, \frac{e^{- i
k_+ x_- - i k_- x_+}}{k^2 + i \epsilon k_0} \, k_+^n \, k_-^m \, = \,
- 2 \, \pi^2 \, \left( \frac{i \, k_T \, \tau}{2 \, x_-} \right)^n \, 
\left( \frac{- i \, k_T \, \tau}{2 \, x_+} \right)^m \, J_{m-n} (k_T \tau),
\ee
where $n$ and $m$ are integers.

\end{document}